\documentclass[fleqn,usenatbib]{mnras}
\usepackage{newtxtext,newtxmath}
\usepackage[T1]{fontenc}
\usepackage{ae,aecompl}
\usepackage{graphicx}	
\usepackage{amsmath}	
\usepackage{amssymb}	
\usepackage{natbib}
\usepackage{titlesec}
\usepackage{setspace}
\usepackage{booktabs}
\usepackage{rotating}
\usepackage{indentfirst}
\usepackage{siunitx}
\usepackage{amsmath} 
\usepackage{todonotes}
\usepackage{array}
\newcolumntype{H}{>{\setbox0=\hbox\bgroup}c<{\egroup}@{}}
\usepackage[flushleft]{threeparttable}
\usepackage{booktabs,caption}
\usepackage{color}
\usepackage[utf8]{inputenc}
\usepackage{hyperref}
\newcommand{\angstrom}{\mbox{\normalfont\AA}}
\newcommand\myrepeat[2]{%
  \begingroup
  \lccode`m=`#2\relax
  \lowercase\expandafter{\romannumeral#1000}%
  \endgroup
}

\newcommand*\mysize{%
   \@setfontsize\mysize{7.5}{9.0}%
}

\newcommand{\surveysize}{23,676 }
\newcommand{\initsize}{40,856 }
\newcommand{\neclipse}{9 }
\newcommand{\nnewpulsator}{41 }
\newcommand{\nknownpulsators}{13 }

\title[Pulsating and Eclipsing WDs from GALEX]{Detections and Constraints on White Dwarf Variability from Time-Series \textit{GALEX} Observations}

\author[Rowan et al.]{
Dominick~M.~Rowan$^{1,2},$
Michael~A.~Tucker$^{1},$
Benjamin~J.~Shappee$^{1},$
and J.~J.~Hermes$^{3}$
\\
$^{1}$Institute for Astronomy, University of Hawaii, 2680 Woodlawn Drive, Honolulu, HI 96822\\
$^{2}$Haverford College, 370 Lancaster Ave, Haverford, PA 19041\\
$^{3}$University of North Carolina, Chapel Hill, NC, 27599
}

\pubyear{\today}

\begin{document}
\label{firstpage}
\pagerange{\pageref{firstpage}--\pageref{lastpage}}
\maketitle

\begin{abstract} 

We search for photometric variability in more than 
23,000 known and candidate white dwarfs, the largest ultraviolet survey compiled for a single study of white dwarfs. We use {\sc gPhoton}, a publicly available calibration/reduction pipeline, to generate time-series photometry of white dwarfs observed by \textit{GALEX}. By implementing a system of weighted metrics, we select sources with variability due to pulsations and eclipses. Although \textit{GALEX} observations have short baselines ($\leq 30$~min), we identify intrinsic variability in sources as faint as \textit{Gaia} $G=20$ mag. With our ranking algorithm, we identify 49 new variable white dwarfs (WDs) in archival \textit{GALEX} observations. We detect \nnewpulsator new pulsators: 37 have hydrogen-dominated atmospheres (DAVs), including one possible massive DAV, and four are helium-dominated pulsators (DBVs). We also detect eight new eclipsing systems; five are new discoveries, and three were previously known spectroscopic binaries. We perform synthetic injections of the light curve of WD 1145+017, a system with known transiting debris, to test our ability to recover similar systems. We find that the $3\sigma$ maximum occurrence rate of WD 1145+017-like transiting objects is $\leq 0.5\%$.
\end{abstract}




\section{Introduction}

More than 97\% of all stars in our Galaxy, including our Sun, will end their lives as white dwarfs (WDs). Not only do WDs offer a lens into degenerate matter physics not replicable on Earth, but they also describe end-stage stellar evolution and planetary systems. By measuring photometric variation in these stars, we can test theoretical models of internal WD structure and energy transfer. 

One way WDs vary is through pulsations. By characterizing the pulsation modes of white dwarf stars, fundamental parameters of the star, including mass, core composition, and internal structure can be determined \citep{Fontaine08,Althaus10}. This technique, known as asteroseismlology, can supplement spectroscopic WD observations, which can constrain $T_{\rm{eff}}$ and $\log g$. Asteroseismology allows us to trace the interior effects of stellar evolution, from the main sequence to the white dwarf stage \citep{Catalan08,Corsico17}. 
 
WD pulsators are grouped by their spectral classification. The majority have hydrogen-dominated atmospheres, known as DAV or ZZ Ceti stars, with temperatures ranging from 10,500--12,500 K \citep{Gianninas11}. Those with helium-dominated atmospheres, DBVs, pulsate at 22,000--29,000 K \citep{Nitta09}. Each type occupies a distinct parameter space in the $\log g - T_{\rm{eff}}$ diagram \citep{Tremblay15}. Within both of these classes, pulsations are driven by partial ionization of the dominant atmospheric element \citep{Winget08}. The brightness variations are generated by global, non-radial gravity ($g$-mode) pulsations that have characteristic timescales typically ranging from $100-1400$ seconds \citep{Koester90,Hermes17}.

Besides pulsations, WDs also often exhibit variability due to eclipses. We expect roughly 29$\pm$8\% of WDs to currently have a binary companion \citep{Toonen17}. There are more than 100 WD+WD binaries known with periods less than a day, although fewer than 10 of those are eclipsing sources \citep{Brown17}. There are also more than 70 eclipsing WD-main sequence binaries that have evolved through a common-envelope phase \citep{Parsons15}.

Finally, it is possible that transiting planets and/or planetary debris can cause variability in a WD light curve. Disintegrating asteroids/planetesimals have been observed around WD 1145+017 \citep{Vanderburg15}. Time-series photometry shows distinct periodicities with broad peaks in the power spectrum, possibly indicating trails of debris following the larger orbiting bodies, which causes deep (up to $>$50\%) transits in the optical. Roughly 1\% of DA WDs have been observed to have an infrared excess from debris disks, indicating the potential for similar detections around sources similar to WD 1145+017 \citep{Farihi16}. WD surveys for transiting planets have been conducted with WASP \citep{Faedi11}, Pan-STARRS1 \citep{Fulton14}, and K2 \citep{vanSluijs18}.  While these surveys yielded no new detections, they placed various upper limits on the occurrence rate for planetary bodies around WDs. 

Characterization of WD variability requires high-cadence photometry with a long baseline. To meet these needs, most WD variability surveys have been conducted in the optical regime \citep[e.g.,][]{Mukadam04}. However, since most WDs have blackbodies that peak in the ultraviolet (UV), it is more promising to search for variability at these wavelengths. Space-based photometry allows for UV observations, though these studies often use much shorter observational baselines. Time-tagged light curves from the \textit{GALEX} space telescope, which observes in two UV bands, was first used for the detection of WD pulsations by \cite{Tucker18}.

In this paper we present variability analysis of \surveysize WDs from archival \textit{GALEX} photometry. Section \S\ref{sourceselection} outlines the source selection criteria for our survey. We develop a ranking method, described in Section \S\ref{ranking}, to identify variability. We report the detection of known variables in Section \S\ref{knownvariables}, new pulsators in Section \S\ref{newpulsators}, and eclipsing binaries in Section \S\ref{neweclipses}. Section \S\ref{recovery_section} describes the injection and recovery of asteroids/planetesimals, as well as upper limits on their occurrence. Finally, we summarize our results in Section \S\ref{conclusion}.

\section{\textit{GALEX} Observations \& Survey Selection} \label{sourceselection}
Photometric observations for WDs in our survey were taken with the \textit{GALEX} space telescope \citep{Morrissey05,Martin05}. \textit{GALEX} observed $\sim77\%~\rm$ of the sky in two ultraviolet bands: the near-ultraviolet (NUV, 1771--2831 \angstrom) and far-ultraviolet (FUV, 1344--1786 \angstrom). 
\textit{GALEX} observations were taken during the night side of each orbit, known as an `eclipse', with each observation lasting up to 30 minutes. Most of our WD sources have been observed by \textit{GALEX} multiple times. We henceforth refer to a single exposure lasting up to 30 minutes as an observation.

\subsection{\sc{gPhoton}}
We implement the publicly available \textit{GALEX} photometry calibration pipeline {\sc gPhoton} \citep[version 1.28.2]{Million16} to check \textit{GALEX} exposure time and produce time-series photometry for each source. We construct light curves with 15-second bins, excluding bins with $<10$ seconds to avoid exposure aliasing. Quality flags output by {\sc gPhoton} help reduce additional instrumental false positives from entering our data. Four quality flags are excluded: mask edge, detector edge, exposure time, and spacecraft recovery \citep{Million16}.

To reduce oversaturation for bright sources, \textit{GALEX} dithers with a 1.5' spiral \citep{Morrissey05,Martin05}. This can represent itself as a periodicity in flux and counts per second, which must be taken into consideration when identifying variability. For an overview of \textit{GALEX}/{\sc gPhoton} instrumental effects and light curve irregularities, see \cite{delaVega18}.

\subsection{Catalog Selection} \label{catalog_selection}
\begin{figure}
\includegraphics[width=\columnwidth]{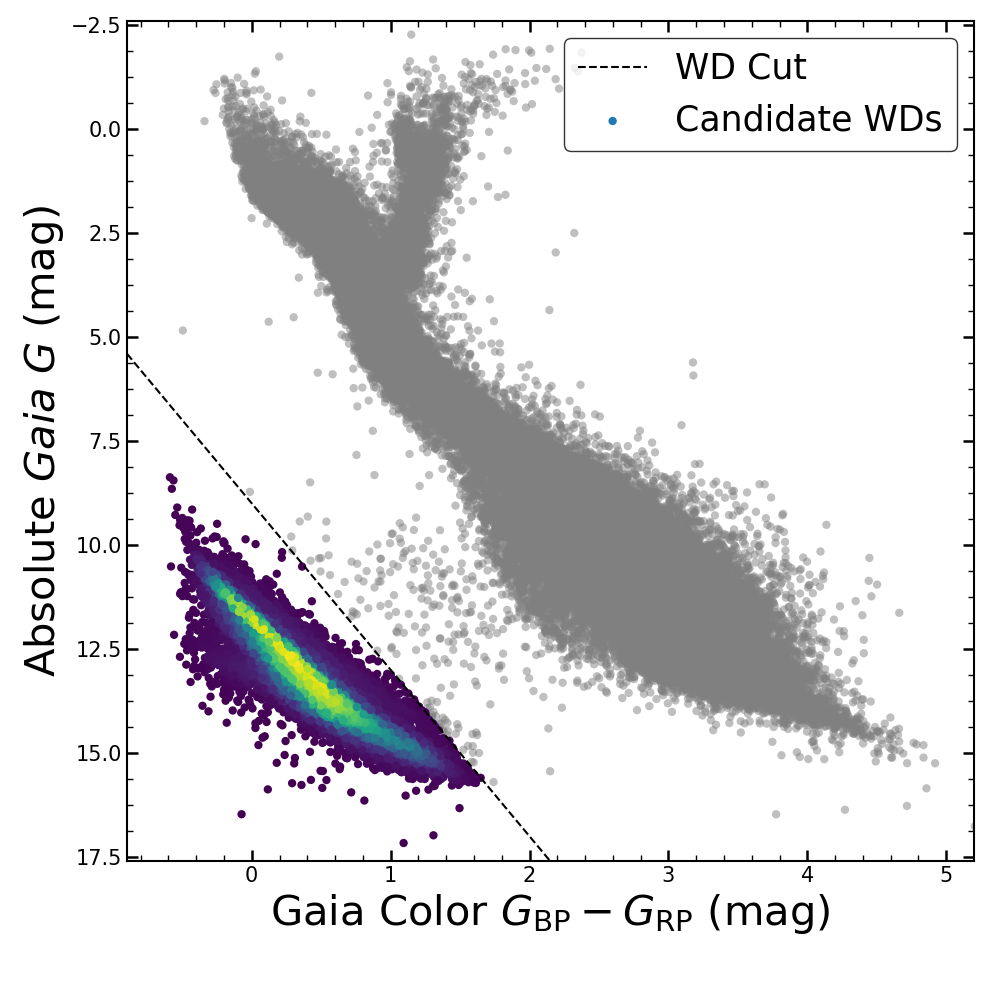}
\caption{We select WDs from the \textit{Gaia} DR2 color magnitude diagram using a linear cut described in Section \S\ref{catalog_selection}. WDs included in our survey are colored by density of points.}
\label{fig:fullcmd}
\end{figure}

The WDs in our survey are taken from 4 catalogs:
\begin{enumerate}
\item Spectroscopically confirmed and high-probability candidate white dwarfs
from the SDSS White Dwarf Catalog \citep[DR8, 10, and 12,][]{Kleinman13,Kepler15,Kepler16}. We exclude any subdwarfs in this catalog.
\item WDs with probability $P_{\rm{WD}} \geq 0.5$ from VST ATLAS \citep{GentileFusillo17}.
\item Sources from Montreal White Dwarf Database \citep[MWDD\footnote{\url{http://dev.montrealwhitedwarfdatabase.org/tables-and-charts.html}}, ][]{Dufour17}.  This database also provides additional information, including binarity and disks, on WDs in other included catalogs. 
\item A cut by color and magnitude in \textit{Gaia} DR2 \citep{GaiaCollaboration18}, selecting sources with the criteria:
\begin{itemize}
\setlength\itemsep{1em}
\item $ \displaystyle G + 5 \log(\frac{\omega}{100}) > 4(G_{\rm{BP}} - G_{\rm{RP}})$
\item $ \displaystyle 1.0+0.015(G_{BP}-G_{RP}) < E_{BP/RP}$
\item $ \displaystyle 1.3+0.06(G_{BP}-G_{RP})^2 > E_{BP/RP}$,

where $G$, $G_{\text{BP}}$, and $G_{\text{RP}}$ are the \textit{Gaia} passbands, $\omega$ is the parallax in milli-arcseconds, and $E_{\text{BP/RP}}$ is the excess factor measured with respect to the \textit{Gaia G} band. 
\end{itemize}
This region is plotted on a color-magnitude diagram in Figure \ref{fig:fullcmd}. The first criteria divides the WD population in color-magnitude space from the non-WD background plotted in gray. The excess factor criteria is a quality cut for background and contamination issues. 
\end{enumerate}

We cross match coordinates between catalogs to remove duplicates, resulting in \initsize sources. The catalog is further reduced by selecting sources with \textit{GALEX} exposure time $>$1000s reported by {\sc gPhoton} to ensure adequate timescales for detectable variability. After applying these cuts and merging the separate catalogs, the final target list consists of \surveysize WDs, the largest photometric ultraviolet WD survey compiled for a single study (previous largest 320, \citealp{Tucker18}).

\section{Identifying Variability} \label{ranking}

To comb through such a large sample, we use a ranking algorithm to prioritize sources more likely to have variability. This system uses four metrics, $c_{LSP}$, $c_{WS}$, $c_{RMS}$, and $c_{EXPT}$. We weight each metric to produce a final rank.

The highest-weighted metric, $c_{LSP}$ with weight $w_{LSP}$, is based on a Lomb-Scargle periodogram \citep[][see \cite{VanderPlas18} for a recent review]{Scargle82, Lomb76} of the light curve. The Baluev Method \citep{Baluev08,VanderPlas18} calculates the false-alarm probability of the five highest amplitudes in each periodogram. Peaks are considered to be statistically significant if the false alarm probability is $<5\%$. This threshold is chosen based on the short baselines of \textit{GALEX} observation, smoothing out high amplitude, narrow peak periodogram features. 

We use a Fourier spectrum of the dither position as a function of time output by {\sc gPhoton} to ensure variability is not due to telescope dither. Peaks overlapping within $8~\rm{seconds}$ of the periodicities in the telescope dither are not considered. For most sources, the dither alias falls at approximately $121~\rm{seconds}$. If the detected period is larger than the exposure time, the metric is multiplied by $1/8$. This consideration helps to reduce some instrumental artifacts, including the jump and slope variations described by \cite{delaVega18}. Since each observation for a source is considered separately, there is no aliasing due to observation cadence. The metric $c_{LSP}$ is calculated as 
\begin{equation}
c_{LSP} = a_{max} / a_{5\%}, 
\end{equation}

\noindent where $a_{max}$ is the amplitude of the highest peak divided by the 5\% false alarm probability level $a_{5\%}$. If there are no peaks with false alarm probability less than $5\%$, $c_{LSP}$ is set to zero.

The second metric is based on the variability index defined by \cite{Welch93}. For GALEX visits with observations in both NUV and FUV, we promote correlated variability in our weighting scheme. Flux residuals are calculated as 
\begin{equation}
\delta_{NUV} = \frac{f_{NUV,i} - \bar{f}_{NUV}}{\sigma_{NUV,i}}
\end{equation}
\begin{equation}
 \delta_{FUV} = \frac{f_{FUV,i} - \bar{f}_{FUV}}{\sigma_{FUV,i}},
\end{equation}
\noindent where $f_{NUV, i}$ and $f_{FUV, i}$ is the flux given on a percentage scale:
\begin{equation}
\text{flux}_{\%} = ((\text{flux}/\text{median}(\text{flux}))-1.0)\times100
\end{equation}
in $i^{th}$ bin for each band. The metric $c_{WS}$, with weight $w_{WS}$, has the functional form
\begin{equation}
c_{WS} = \frac{1}{n}\sqrt{\frac{n}{n-1}}\sum_{i=1}^n \delta_{NUV,i} \delta_{FUV,i}
\end{equation}
where $n$ is the number of discrete photometric measurements in the light curve. For random observations, the $i^{th}$ $\delta_{NUV}$ and $\delta_{FUV}$ values will be uncorrelated, averaging out to 0 for the metric \citep{Welch93}. For light curves without concurrent coverage in both bands we set the $\delta_{other}$, either $\delta_{FUV}$ or $\delta_{NUV}$, to be 1 in the calculation. Since telescope dither is consistent across both bands, this is an additional consideration to reduce false positives. If there is a prominent periodicity (false alarm probability $<25\%$) in the light curve data corresponding to the dither period, we exclude this metric, setting $c_{WS}$ to be zero for the observation. This metric is similar to $c_{RMS}$. We include both in order to promote correlated variability without penalizing single-band observations.

The third-highest-weighted metric, $c_{RMS}$ with weight $w_{RMS}$, considers how the root mean square (RMS) in magnitude ($\sigma_{RMS}$) for a source compares to sources of similar magnitudes. Sources with high amplitudes of variability are expected to have larger scatter in magnitude, as compared to non-varying WDs of a similar magnitude. WDs are binned by NUV and FUV magnitude, with bin sizes 0.1 mag. The $\sigma_{RMS}$ of each source is then compared to the median $\sigma_{RMS}$ of sources in the same magnitude bin. The metric is calculated as $c_{RMS} = RMS / \rm{median}(RMS)$. Many sources have especially high RMS values due to instrumental effects. To reduce this impact on our ranking, we cap the RMS metric at the $95^{\rm{th}}$ percentile, which reduces the false positive rate while still retaining the effectiveness of the metric. This metric is especially helpful in the identification of eclipses, where the decrease in flux is consistent with zero flux. 

The final metric is based on exposure time and data-quality flags. Variability is more likely to be detected in sources with longer timescales of observation. Therefore, we use the metric $c_{EXP}$ with weight $w_{EXP}$ to take exposure time into consideration. However, in order to include consideration of removed flagged data, the metric excludes time in flagged bins, computed as $c_{EXP} = t_{\rm{total}} - t_{\rm{flagged}}$. 

For each \textit{GALEX} observation, the rank R is calculated as the sum of the metrics:
\begin{equation}
\begin{split}
    \mathrm{R} = & w_{LSP}c_{LSP} + w_{WS}c_{WS} +  w_{RMS}c_{RMS} + w_{EXP}c_{EXP} 
\end{split}
\end{equation}

Since most of our WDs have multiple distinct \textit{GALEX} observations, each lasting up to 30 minutes, the final rank for a given WD is the max R value among all observations. This is useful for sifting through sources with long-term or aperiodic variability. The purpose of our ranking system is to quickly identify variability rather than to provide an absolute methodology to compare variable sources. Therefore, weights are adjusted manually using random subsets of our sample to ensure known variable WDs are recovered successfully. The final weights used are given in Table \ref{tab:weights}. After tuning weights on a smaller scale, we apply the ranking algorithm to the entire sample, producing a list with higher ranked sources more likely to display astrophysical variability.

Even after implementing our ranking system, we do not completely remove false positives from the data. Specifically, there are sources with periodicities due to telescope dither that occasionally have harmonics at higher frequencies. Some FUV observations show a recurring long-term variability that does not correlate with NUV trends. These `jump variations', described in \cite{delaVega18}, are likely due to instrumental effects with the FUV filter \citep{Morrissey06}.

\begin{figure*}
    \centering
    \includegraphics[width=\linewidth]{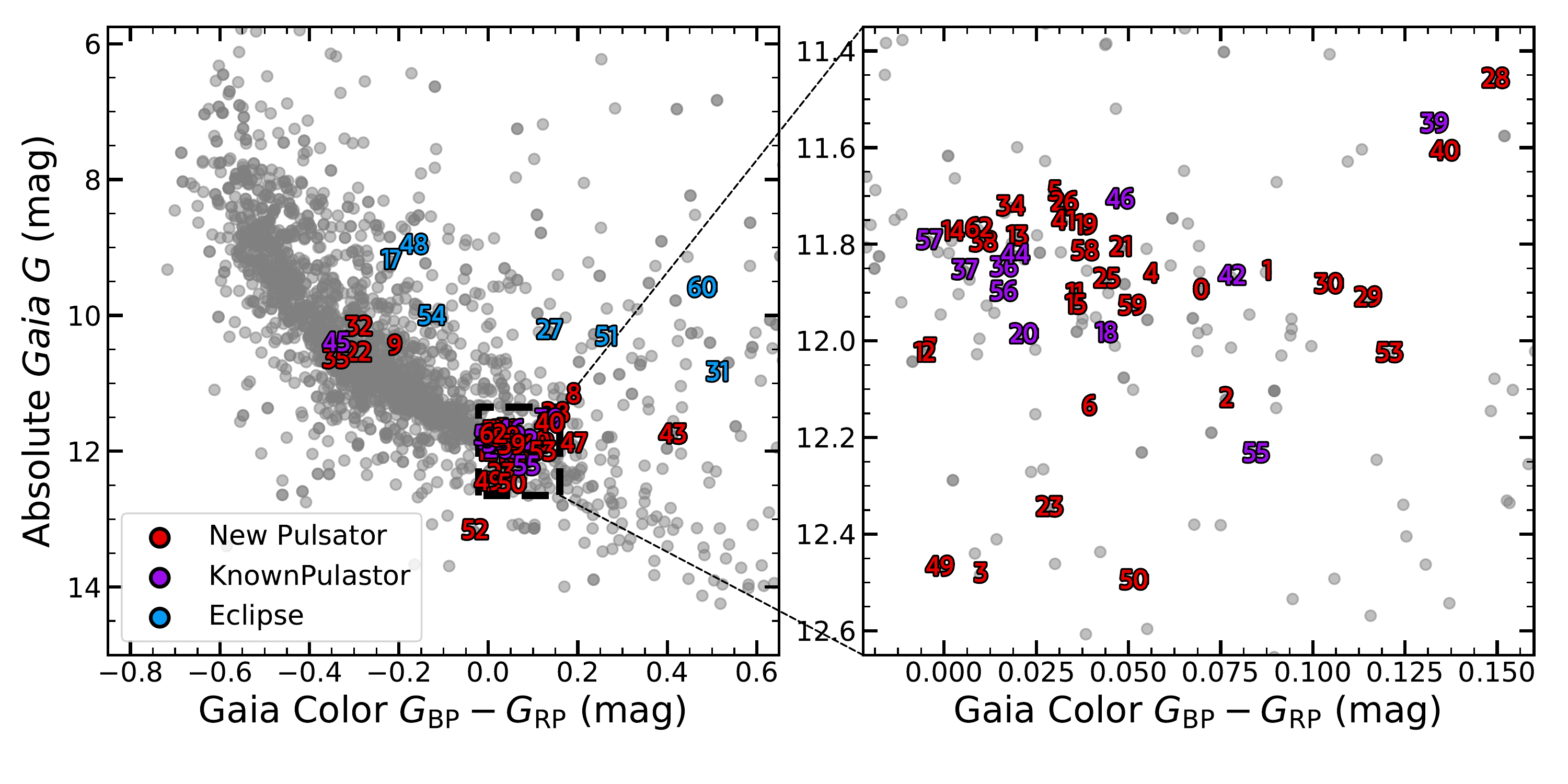}
    \caption{Left: \textit{Gaia} color-magnitude diagram for the top 3000 ranked sources in our survey. \nknownpulsators Known pulsators, \nnewpulsator new pulsators, and \neclipse eclipsing sources are labeled with their identification number, colored in purple, red, and blue, respectively. We observe two separate populations - DAV stars at $G_{\rm{BP}}-G_{\rm{RP}}\approx 0.1$, M$_{G}\approx12$ and DBV pulsators at $G_{\rm{BP}}-G_{\rm{RP}}\approx -0.3$, M$_{G}\approx 10.5$. Eclipsing sources are overluminous compared to typical WDs at their respective $G_{\rm{BP}}-G_{\rm{RP}}$ colors. Right: Close-in view of instability region for DAV stars.} 
    \label{fig:cmd}
\end{figure*}

We inspect the top $\sim10\%$ of ranked sources manually to characterize the type of variability and to remove false positives. After identifying sources with intrinsic variability, we divide our sample into three separate categories: known pulsators, new pulsators, and eclipsing sources. Out of the total \surveysize WDs, we detect 41 new pulsators, \nknownpulsators known pulsators, 8 new eclipsing WDs, and 1 known eclipsing WD. We plot all 63 variable WDs detected on a \textit{GAIA} CMD in Figure \ref{fig:cmd}.

\setlength{\tabcolsep}{4pt}
\begin{table}
\caption{Weights used in the ranking algorithm described in \S\ref{ranking}.}
\label{tab:weights}
\begin{tabular}{llr}
\toprule
                   Metric &        Abbreviation &  Weight \\
\midrule
 Lomb-Scargle Periodogram &           $w_{LSP}$ &        0.5714 \\
      Welch-Stetson Index &            $w_{WS}$ &        0.1714 \\
           RMS			  &  $w_{RMS}$ &                 0.1429 \\
            Exposure (ks) &          $w_{EXPT}$ &        0.1143 \\
\bottomrule
\end{tabular}

\end{table}

\begin{figure}
	\centering
    \includegraphics[width=\columnwidth]{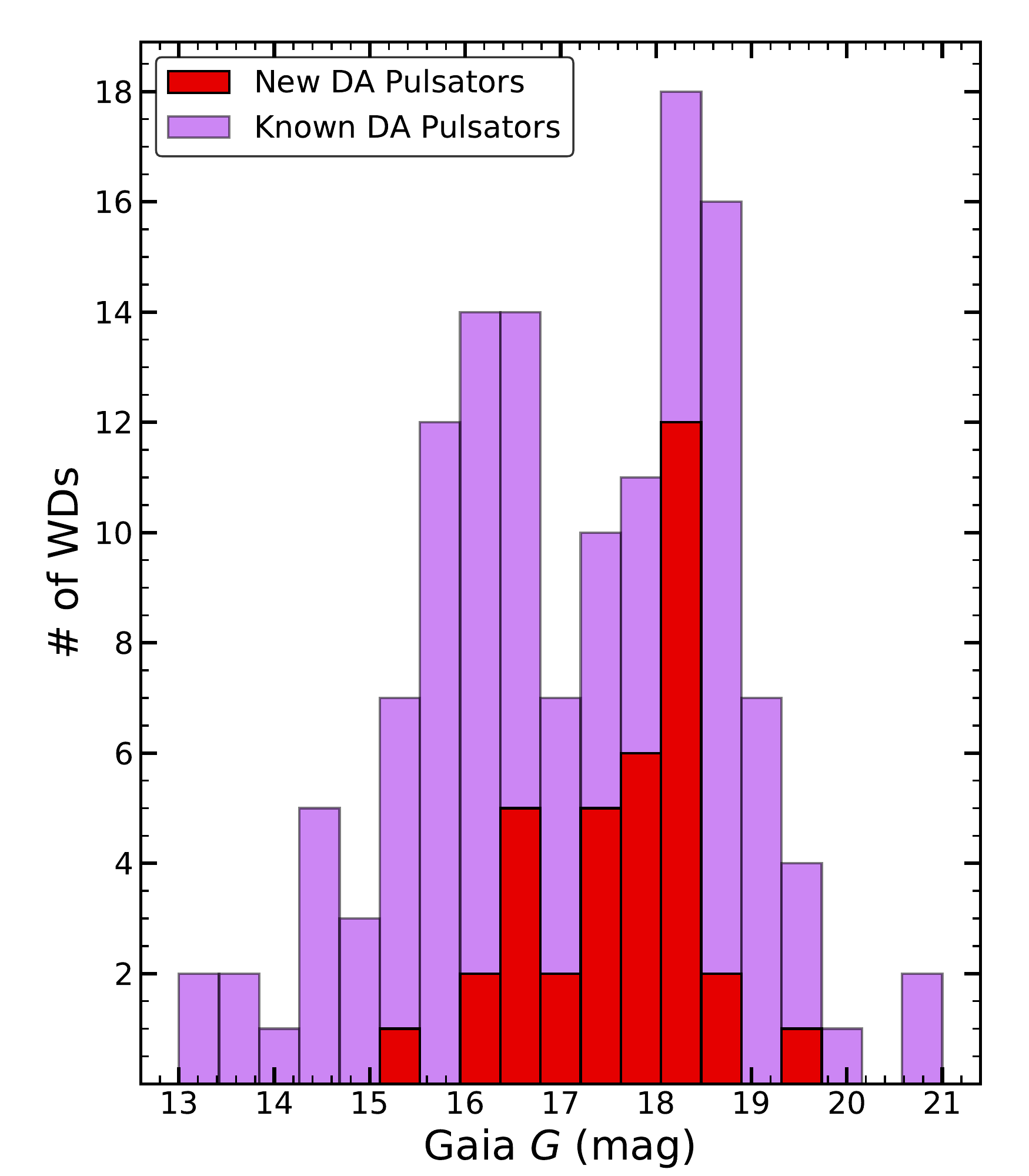}
    \caption{Histogram of apparent magnitudes for new DAV pulsators detected compared to known DAV pulsators in Table 4 of \protect\cite{Bognar16}.}
    \label{fig:maghist}
\end{figure}

\section{Known Variables} \label{knownvariables}
In addition to the 49 new variables detected, there are \nknownpulsators known pulsators and one known eclipsing WD in our survey. Identification information for all sources is provided in Table \ref{tab:allsources}, and discovery references are given in Table \ref{tab:knowntable}. UV light curves for these sources are plotted in Appendix \ref{appendixLC}. 

\setlength{\tabcolsep}{4pt}
\begin{table*}
\vskip-.2 in 
\fontsize{6.}{7.}
\begin{threeparttable}
\caption{Information on all detected variable WDs. ID numbers are assigned for each variable in this work. Variability period, NUV amplitude, and FUV amplitude are estimated by fitting sinusoids to the light curves. References for known variables are given in Table \ref{tab:knowntable}.}
\label{tab:allsources}
\begin{tabular}{lrrrrlllll}
\toprule
                     Source &  ID\# &         RA &      Dec. &  $Gaia$ $G$ &  Classification &       Spec. Type &       Period &    NUV Amp &    FUV Amp \\
& & (deg) & (deg) & (mag) & & & (s) & (\%) & (\%) \\ 
\midrule
  GALEX 2417766833348155164 &     0 &    7.44064 &  14.70402 &       18.20 &    New Pulsator &              DAV &   898$\pm$44 &   20$\pm$2 &   52$\pm$7 \\
  WD J002959.22$+145812.97$ &     1 &    7.49674 &  14.97027 &       17.49 &    New Pulsator &  DAV$^{\dagger}$ &   948$\pm$47 &    9$\pm$2 &   41$\pm$5 \\
  WD J003116.51$+474828.39$ &     2 &    7.81881 &  47.80789 &       18.36 &    New Pulsator &  DAV$^{\dagger}$ &   896$\pm$41 &   18$\pm$3 &   47$\pm$9 \\
  WD J004154.68$-030802.71$ &     3 &   10.47781 &  -3.13409 &       18.07 &    New Pulsator &  DAV$^{\dagger}$ &   865$\pm$55 &    5$\pm$4 &  17$\pm$11 \\
   WD J010025.6$+421840.22$ &     4 &   15.10665 &  42.31117 &       16.58 &    New Pulsator &  DAV$^{\dagger}$ &   574$\pm$27 &    8$\pm$1 &   26$\pm$8 \\
  WD J010528.83$+020500.96$ &     5 &   16.37012 &   2.08360 &       16.73 &    New Pulsator &  DAV$^{\dagger}$ &   710$\pm$14 &   12$\pm$1 &   34$\pm$3 \\
  WD J010539.17$+321846.53$ &     6 &   16.41320 &  32.31292 &       18.07 &    New Pulsator &  DAV$^{\dagger}$ &   407$\pm$10 &    9$\pm$2 &   35$\pm$7 \\
           KUV 01595$-$1109 &     7 &   30.49558 & -10.91064 &       16.89 &    New Pulsator &              DAV &   687$\pm$15 &   14$\pm$1 &         -- \\
  WD J022941.29$-063842.89$ &     8 &   37.42204 &  -6.64525 &       18.23 &    New Pulsator &  DAV$^{\dagger}$ &   754$\pm$16 &   23$\pm$2 &   77$\pm$8 \\
  WD J025121.71$-125244.85$ &     9 &   42.84045 & -12.87913 &       18.23 &    New Pulsator &  DBV$^{\dagger}$ &   944$\pm$72 &    7$\pm$2 &   11$\pm$3 \\
  2MASS J03030835$+$0054438 &    10 &   45.78483 &   0.91225 &       17.38 &         Eclipse &             DC+M &           -- &         -- &         -- \\
  WD J030648.49$-172332.19$ &    11 &   46.70206 & -17.39228 &       16.70 &    New Pulsator &  DAV$^{\dagger}$ &   794$\pm$15 &   17$\pm$1 &   72$\pm$4 \\
  WD J053212.77$-432006.05$ &    12 &   83.05323 & -43.33501 &       18.19 &    New Pulsator &  DAV$^{\dagger}$ &   451$\pm$15 &   11$\pm$6 &         -- \\
   WD J080609.2$+111230.83$ &    13 &  121.53832 &  11.20856 &       18.11 &    New Pulsator &  DAV$^{\dagger}$ &    391$\pm$7 &    8$\pm$6 &  30$\pm$13 \\
  WD J084055.73$+130329.38$ &    14 &  130.23221 &  13.05816 &       17.22 &    New Pulsator &  DAV$^{\dagger}$ &   490$\pm$16 &   12$\pm$2 &   34$\pm$5 \\
                    US 1639 &    15 &  131.72060 &  44.44405 &       18.19 &    New Pulsator &              DAV &   663$\pm$33 &   16$\pm$3 &  58$\pm$10 \\
            WD J0855$+$0635 &    16 &  133.78021 &   6.59446 &       17.32 &  Known Pulsator &              DAV &   895$\pm$24 &   19$\pm$2 &   51$\pm$6 \\
  WD J090023.05$+434813.45$ &    17 &  135.09604 &  43.80374 &       18.38 &         Eclipse &             D?+? &           -- &         -- &         -- \\
            WD J0906$-$0024 &    18 &  136.60098 &  -0.40776 &       17.87 &  Known Pulsator &              DAV &   747$\pm$27 &    8$\pm$6 &  14$\pm$18 \\
  WD J093250.51$+554315.68$ &    19 &  143.21045 &  55.72102 &       17.69 &    New Pulsator &  DAV$^{\dagger}$ &   510$\pm$17 &   10$\pm$2 &   35$\pm$5 \\
              WD 0938$+$577 &    20 &  145.55490 &  57.56188 &       17.49 &  Known Pulsator &              DAV &   608$\pm$16 &   13$\pm$3 &         -- \\
 SDSS J094851.43$+$512448.0 &    21 &  147.21425 &  51.41336 &       18.57 &    New Pulsator &              DAV &   947$\pm$63 &  14$\pm$10 &  32$\pm$30 \\
 SDSS J102106.69$+$082724.8 &    22 &  155.27769 &   8.45684 &       17.82 &    New Pulsator &              DBV &   471$\pm$11 &    8$\pm$1 &   11$\pm$2 \\
 SDSS J103642.25$+$211527.9 &    23 &  159.17583 &  21.25773 &       17.60 &    New Pulsator &              DAV &   679$\pm$41 &    7$\pm$3 &   23$\pm$5 \\
                    CBS 130 &    24 &  162.68750 &  33.26500 &       16.56 &    New Pulsator &              DAV &    484$\pm$9 &    7$\pm$1 &   30$\pm$4 \\
  WD J105046.04$+331546.22$ &    25 &  162.69184 &  33.26284 &       16.56 &    New Pulsator &  DAV$^{\dagger}$ &    484$\pm$9 &    7$\pm$1 &   31$\pm$4 \\
 SDSS J110505.94$+$583103.0 &    26 &  166.27392 &  58.51735 &       17.94 &    New Pulsator &              DAV &   825$\pm$20 &   19$\pm$2 &   81$\pm$7 \\
              WD 1104$+$656 &    27 &  167.02229 &  65.36986 &       18.53 &         Eclipse &             DA+M &           -- &         -- &         -- \\
  WD J115057.43$-055306.58$ &    28 &  177.73930 &  -5.88516 &       17.44 &    New Pulsator &  DAV$^{\dagger}$ &    424$\pm$9 &    6$\pm$2 &  18$\pm$10 \\
    USNO A2.0 1350-08049341 &    29 &  180.78821 &  45.75564 &       18.57 &    New Pulsator &              DAV &   835$\pm$30 &   23$\pm$3 &         -- \\
   WD J122155.73$+050621.6$ &    30 &  185.48222 &   5.10600 &       17.90 &    New Pulsator &  DAV$^{\dagger}$ &  1061$\pm$51 &   19$\pm$2 &         -- \\
  2MASS J12233961$-$0056311 &    31 &  185.91504 &  -0.94200 &       17.84 &         Eclipse &             DA+M &           -- &         -- &         -- \\
 SDSS J123654.96$+$170918.7 &    32 &  189.22881 &  17.15514 &       18.17 &    New Pulsator &              DBV &   698$\pm$28 &    9$\pm$1 &         -- \\
 SDSS J124759.03$+$110703.0 &    33 &  191.99596 &  11.11750 &       19.35 &    New Pulsator &              DAV &   672$\pm$46 &  29$\pm$10 &  61$\pm$50 \\
  WD J124804.04$+282103.46$ &    34 &  192.01683 &  28.35096 &       18.04 &    New Pulsator &  DAV$^{\dagger}$ &   678$\pm$20 &   13$\pm$4 &         -- \\
   WD J132952.63$+392150.8$ &    35 &  202.46928 &  39.36411 &       18.01 &    New Pulsator &  DBV$^{\dagger}$ &   727$\pm$34 &    7$\pm$1 &         -- \\
            WD J1355$+$5454 &    36 &  208.87929 &  54.90125 &       18.67 &  Known Pulsator &              DAV &    158$\pm$4 &   14$\pm$3 &   15$\pm$7 \\
                  V* IU Vir &    37 &  210.98747 & -15.02017 &       15.75 &  Known Pulsator &              DAV &    586$\pm$8 &    9$\pm$1 &   37$\pm$2 \\
              WD 1452$+$600 &    38 &  223.34756 &  59.84885 &       17.18 &    New Pulsator &              DAV &   530$\pm$35 &    5$\pm$2 &   26$\pm$8 \\
            WD J1502$-$0001 &    39 &  225.52925 &  -0.02975 &       18.72 &  Known Pulsator &              DAV &   645$\pm$35 &   11$\pm$4 &  80$\pm$13 \\
   WD J150626.13$+063845.0$ &    40 &  226.60887 &   6.64583 &       16.60 &    New Pulsator &  DAV$^{\dagger}$ &    635$\pm$8 &   10$\pm$2 &  21$\pm$16 \\
  WD J150739.31$+074828.54$ &    41 &  226.91378 &   7.80793 &       18.21 &    New Pulsator &  DAV$^{\dagger}$ &   583$\pm$23 &   14$\pm$2 &   32$\pm$6 \\
            WD J1617$+$4324 &    42 &  244.40697 &  43.41215 &       18.44 &  Known Pulsator &              DAV &   645$\pm$16 &    9$\pm$9 &         -- \\
  WD J162724.72$+392027.25$ &    43 &  246.85299 &  39.34090 &       16.23 &    New Pulsator &  DAV$^{\dagger}$ &    268$\pm$4 &    5$\pm$1 &   18$\pm$2 \\
             HS 1625$+$1231 &    44 &  247.05510 &  12.41426 &       16.27 &  Known Pulsator &              DAV &   974$\pm$12 &   16$\pm$1 &   55$\pm$3 \\
                     GD 358 &    45 &  251.82582 &  32.47590 &       13.59 &  Known Pulsator &              DAV &   874$\pm$15 &    7$\pm$0 &    8$\pm$0 \\
            WD J1700$+$3549 &    46 &  255.23032 &  35.83057 &       17.36 &  Known Pulsator &              DAV &   924$\pm$31 &   17$\pm$3 &  46$\pm$26 \\
  WD J173351.44$+341012.96$ &    47 &  263.46433 &  34.17027 &       16.35 &    New Pulsator &  DAV$^{\dagger}$ &   793$\pm$31 &   10$\pm$1 &         -- \\
   WD J193639.33$-524600.1$ &    48 &  294.16388 & -52.76669 &       18.02 &         Eclipse &             D?+? &           -- &         -- &         -- \\
  WD J202838.13$-060842.11$ &    49 &  307.15889 &  -6.14503 &       15.22 &    New Pulsator &  DAV$^{\dagger}$ &   773$\pm$26 &    4$\pm$1 &         -- \\
  WD J204127.11$-041724.22$ &    50 &  310.36297 &  -4.29006 &       18.31 &    New Pulsator &  DAV$^{\dagger}$ &   527$\pm$34 &   17$\pm$4 &  29$\pm$10 \\
 SDSS J212232.58$-$061839.7 &    51 &  320.63575 &  -6.31103 &       19.44 &         Eclipse &              DAV &           -- &         -- &         -- \\
  WD J212402.03$-600100.05$ &    52 &  321.00847 & -60.01668 &       17.97 &    New Pulsator &  DAV$^{\dagger}$ &    357$\pm$3 &   16$\pm$2 &   60$\pm$5 \\
   WD J215321.8$+044019.92$ &    53 &  328.34083 &   4.67220 &       18.19 &    New Pulsator &  DAV$^{\dagger}$ &   805$\pm$43 &    9$\pm$2 &   51$\pm$8 \\
 SDSS J220823.66$-$011534.1 &    54 &  332.09858 &  -1.25944 &       18.60 &         Eclipse &             DA+? &           -- &         -- &         -- \\
            WD J2209$-$0919 &    55 &  332.31600 &  -9.32847 &       18.61 &  Known Pulsator &              DAV &   720$\pm$44 &   14$\pm$3 &  45$\pm$10 \\
             HE 2246$-$0658 &    56 &  342.16694 &  -6.71254 &       16.94 &  Known Pulsator &              DAV &   659$\pm$12 &    9$\pm$4 &   39$\pm$8 \\
                     GD 244 &    57 &  344.19314 &  12.88068 &       15.78 &  Known Pulsator &              DAV &    272$\pm$3 &    4$\pm$1 &   17$\pm$2 \\
   WD J231536.9$+192448.81$ &    58 &  348.90375 &  19.41356 &       17.99 &    New Pulsator &  DAV$^{\dagger}$ &   741$\pm$43 &   15$\pm$2 &         -- \\
  WD J231641.17$-315352.74$ &    59 &  349.17152 & -31.89798 &       18.29 &    New Pulsator &  DAV$^{\dagger}$ &    373$\pm$7 &    4$\pm$2 &  21$\pm$17 \\
  GALEX 2667197544394657586 &    60 &  355.50938 &  -3.47822 &       19.37 &         Eclipse &             DA+M &           -- &         -- &         -- \\
 SDSS J234829.09$-$092500.9 &    61 &  357.12121 &  -9.41693 &       19.64 &         Eclipse &             DA+? &           -- &         -- &         -- \\
   WD J235010.36$+201913.9$ &    62 &  357.54317 &  20.32053 &       17.41 &    New Pulsator &  DAV$^{\dagger}$ &    333$\pm$8 &    7$\pm$2 &   29$\pm$4 \\
\bottomrule
\end{tabular}

\begin{tablenotes}
\fontsize{6.5}{7.8}
\item $^{\dagger}$ Spectral type classification based of position on \textit{Gaia} CMD.
\end{tablenotes}
\end{threeparttable}
\end{table*}

\setlength{\tabcolsep}{4pt}
\begin{table}
\caption{References for known WD pulsators and binaries detected.}
\label{tab:knowntable}
\begin{tabular}{lrl}
\toprule
                    Source &  ID\# &             Reference \\
\midrule
 2MASS J03030835$+$0054438 &    10 &   \cite{Eisenstein06} \\
           WD J0855$+$0635 &    16 &  \cite{Castanheira07} \\
           WD J0906$-$0024 &    18 &      \cite{Mukadam04} \\
             WD 0938$+$577 &    20 &      \cite{Mukadam04} \\
             WD 1104$+$656 &    27 &    \cite{Silvestri06} \\
 2MASS J12233961$-$0056311 &    31 &   \cite{Eisenstein06} \\
           WD J1355$+$5454 &    36 &     \cite{Mullally05} \\
                 V* IU Vir &    37 &       \cite{Stobie95} \\
           WD J1502$-$0001 &    39 &      \cite{Mukadam04} \\
           WD J1617$+$4324 &    42 &      \cite{Mukadam04} \\
            HS 1625$+$1231 &    44 &         \cite{Voss06} \\
                    GD 358 &    45 &        \cite{Kotak03} \\
           WD J1700$+$3549 &    46 &      \cite{Mukadam04} \\
           WD J2209$-$0919 &    55 &  \cite{Castanheira10} \\
            HE 2246$-$0658 &    56 &       \cite{Tucker18} \\
                    GD 244 &    57 &     \cite{Fontaine01} \\
 GALEX 2667197544394657586 &    60 &       \cite{Kepler15} \\
\bottomrule
\end{tabular}

\end{table}

\begin{figure*}
    \centering
    \includegraphics[width=\linewidth, height=10in, keepaspectratio]{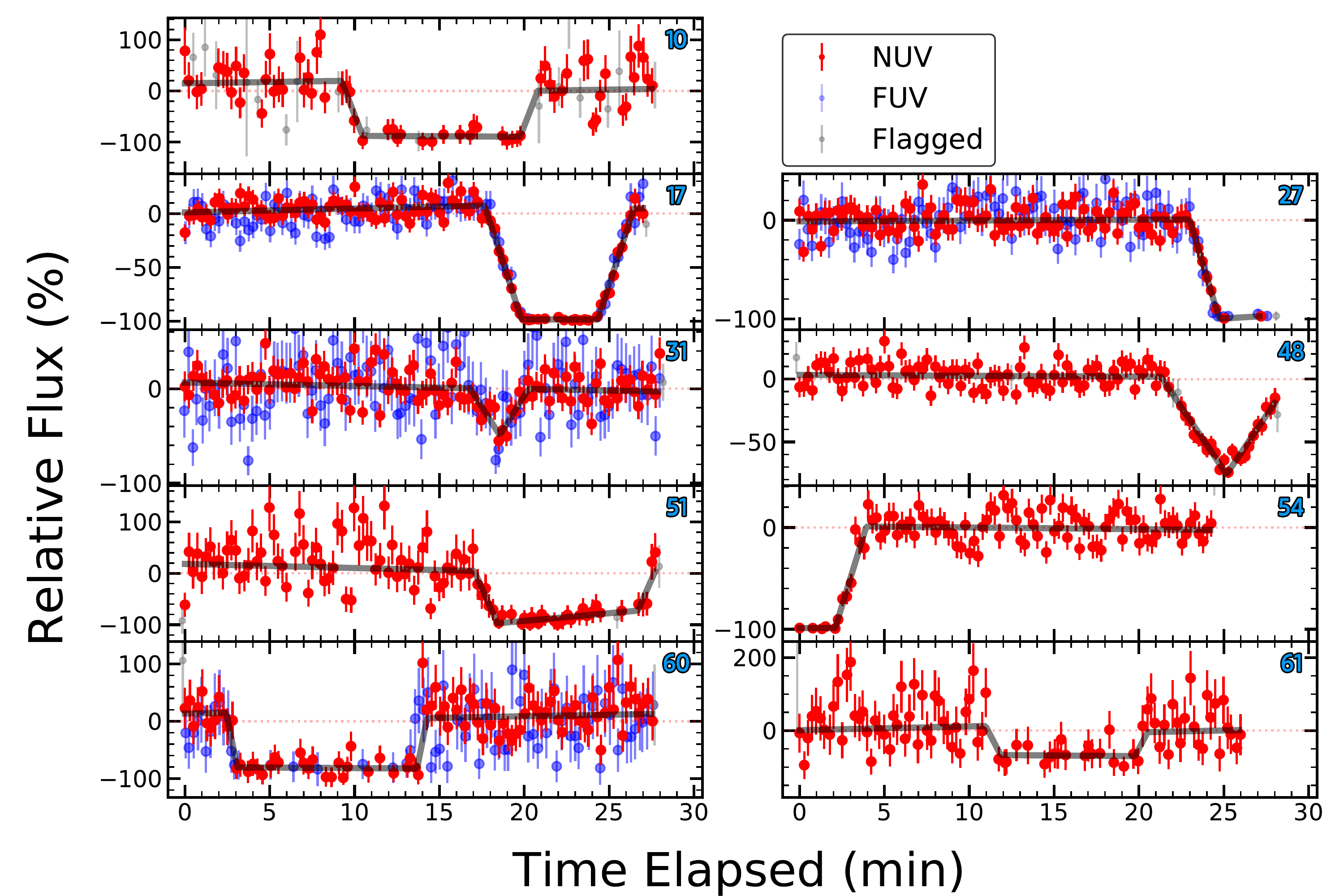}
    \caption{Light curves for the nine eclipsing sources in our sample. WD ID\# from Table \protect\ref{tab:allsources} is given in the top right of each panel. When available, FUV data is overplotted in blue and flagged points are plotted in gray. We fit simple trapezoidal eclipse models (black) to estimate eclipse duration and center, which are given in Table \ref{tab:eclipseinfo}.}
    \label{fig:EclipseTower}
\end{figure*}

One known eclipsing binary is included in our sample, US 3566 (2MASS J03030835$+$0054438, ID\# 10), which was spectroscopically characterized as a WD and M4-dwarf (DC+dM4) binary system with an orbital period of $3.2$ hours \citep{Eisenstein06,Pyrzas09}. \cite{Debes12} classified the system as a post-common-envelope binary with infrared excess, at $3\mu$m using WISE infrared observations. Follow-up observations show that the white dwarf in the system is strongly magnetic (8 MG) and that this IR excess is due to cyclotron emission; this pre-cataclysmic variable will likely evolve into an intermediate polar in roughly 1 Gyr \citep{Parsons2013}.

There are five \textit{GALEX} observations of US 3566, two of which show a partial-duration and a full-duration eclipse, plotted in Figure \ref{fig:EclipseTower}. The gap in \textit{GALEX} observations prevents further refinement of the US 3566 orbital parameters presented in \cite{Eisenstein06}.

\section{New Pulsators} \label{newpulsators}
Out of the 63 variables detected, we find \nnewpulsator new WD pulsators in our sample which are outlined in Table~\ref{tab:allsources}. Light curves for the highest ranked visit of each source are plotted in Appendix~\ref{appendixLC}. The new pulsators detected have a median \textit{Gaia} $G=18.0$ mag. Figure \ref{fig:maghist} compares the distribution of new pulsator magnitudes to known DA pulsators listed in Table 4 of \cite{Bognar16}. Many bright sources that might be traditional candidates for variability are overexposed in the \textit{GALEX} aperture, leading to dither effects and high flag ratios of $>25\%$. There are a number of known pulsators in which we do not detect variation due to overexposure and other instrumental effects.

One way pulsators can be distinguished spectroscopically is by consideration of their $\log g$ and $T_{\rm{eff}}$, given their location in an instability strip \citep[Figure 35, ][]{Gianninas11}. We construct a similar parameter space in a \textit{Gaia} DR2 color-magnitude diagram (CMD) in Figure \ref{fig:cmd}. Both the DAV and DBV instability strips occupy distinct locations on the \text{Gaia} CMD. We find most pulsators fall into the DAV instability strip, labeled in Figure \ref{fig:cmd}; this is expected since more than 80 percent of white dwarfs in magnitude-limited samples have hydrogen-dominated atmospheres \citep{Kleinman13}. These WDs could also be photometrically variable due to surface spots, as in hot DQVs \citep{Kurtis16}. However, the pulse shapes and \textit{Gaia} CMD positions suggest the variations are due to pulsations and not rotationally driven spots.

We include spectra for two new pulsators in Appendix \ref{appendixSpectra} showing expected H/He spectral features. We emphasize that we did not search for variability based on CMD location. Thus, sources falling in known instability strips provides additional confidence in our method and their classification. 

\subsection{WD J212402.03$-$600100.05: A New Massive Pulsator}

This pulsator, ID \#52, has an absolute \textit{Gaia G} magnitude roughly $1$ mag fainter than the rest of the DAVs (Figure~\ref{fig:cmd}). This is likely a more massive WD pulsator, similar to GD 518 \citep{Hermes13b}, since WDs decrease in radius with increasing mass. Hydrogen-atmosphere model fits of the \textit{Gaia} CMD position suggest this is a roughly $12510\pm750$ K WD with a mass of $1.16\pm0.04$ $\text{M}_{\odot}$ \citep{GentileFusillo18}. Massive WD pulsators are of particular interest since the cores can be oxygen/neon or crystallized. Additional follow up photometric and spectroscopic observations and classification will be presented in an upcoming paper.

\subsection{WD J162724.72$+$392027.25}
One DA pulsator, ID \#43, is $\sim 0.2$ mag redder than the rest of the population. Based on the light curves presented in Appendix \ref{fig:LCappendix1}, we are confident the variability detected is due to pulsations. Inspecting the Pan-STARRS images \citep{Chambers16,Flewelling16}\footnote{\url{http://ps1images.stsci.edu/cgi-bin/ps1cutouts}} of this WD, there is a relatively bright red star ($r\sim13.6$ mag, $g-r\sim0.7$ mag) approximately $2^{\prime\prime}$ away from the WD. Considering the \textit{Gaia} point spread functions extend out to $1$-$2^{\prime\prime}$ \citep{Fabricius16}, the redward shift of ID \#43 compared to the rest of the DAV population is likely due to blending of these sources. Thus, we still consider this object a new DAV pulsator despite its anomalous location in Figure \ref{fig:cmd}.

\subsection{DB Pulsators}
The CMD is also helpful in identifying non-DAV pulsators our variability sample. One of the known pulsators detected, GD 358, (ID \#45), is a known DBV pulsator \citep{Kotak03}. We find two new pulsators that are spectroscopically known to be DB type WDs, ID\#22 and \#32, as well as two others with unknown spectral types, ID\#9 and \#35, that occupy a similar parameter space as GD 358 on the CMD. Therefore, we classify these four new pulsators as new DB pulsators (DBVs).

\section{New Eclipsing WDs} \label{neweclipses}

We detect eight new eclipsing binaries in addition to the pulsators. These WDs are overluminous in the Gaia CMD (Figure \ref{fig:cmd}), suggesting the excess flux is from their companions. For the three sources that are known spectroscopic binary systems, we present the first evidence of eclipses \citep[ID\#27, 31, and 60,][]{Silvestri06,Eisenstein06,Kepler15}. The light curves for all eclipsing sources are plotted in Figure \ref{fig:EclipseTower}.

Two of the eclipses  detected (ID\#31, 48) are grazing eclipses, where the drop in flux is not consistent with $100\%$ decrease in flux. There are also two eclipses that are partially observed (ID\#27, 54). Though we are limited by the short timescales of \textit{GALEX} observations, we construct simple trapezoidal eclipse models to estimate eclipse duration and center. The results of these fits are reported in Table \ref{tab:eclipseinfo}.

\begingroup
\renewcommand{\arraystretch}{1.2}
\setlength{\tabcolsep}{4pt}
\begin{table}
\caption{Data on eclipsing WDs in our survey. Parameters are derived using trapezoidal transit fits. Eclipse center is given in minutes from observation start.}
\label{tab:eclipseinfo}
\begin{tabular}{rlcc}
\toprule
 ID &       Obs Start &            Eclipse Center &                  Duration \\
 & (TDB JD) & (min from start) & (min) \\ 
\midrule
 10 &  2455164.301501 &  $15.01^{+ 0.21}_{-0.42}$ &  $11.56^{+ 0.96}_{-0.82}$ \\
 17 &  2453436.107338 &  $22.06^{+ 0.06}_{-0.06}$ &   $8.80^{+ 0.17}_{-0.20}$ \\
 27 &  2453027.340377 &                        -- &                $\gtrsim4$ \\
 31 &  2454946.159964 &  $18.54^{+ 0.13}_{-0.08}$ &   $3.45^{+ 0.84}_{-0.49}$ \\
 48 &  2455387.297459 &  $25.16^{+ 0.06}_{-0.07}$ &   $6.74^{+ 0.51}_{-0.39}$ \\
 51 &  2455039.073697 &  $23.29^{+ 0.60}_{-0.30}$ &  $12.48^{+ 1.19}_{-0.60}$ \\
 54 &  2453254.882559 &                        -- &                $\gtrsim3$ \\
 60 &  2454716.101505 &   $8.35^{+ 0.10}_{-0.09}$ &  $11.83^{+ 0.41}_{-0.29}$ \\
 61 &  2455468.906436 &  $15.73^{+ 0.15}_{-0.18}$ &   $9.55^{+ 0.38}_{-0.34}$ \\
\bottomrule
\end{tabular}

\end{table}
\endgroup

\begin{figure*}
\includegraphics[width=\linewidth]{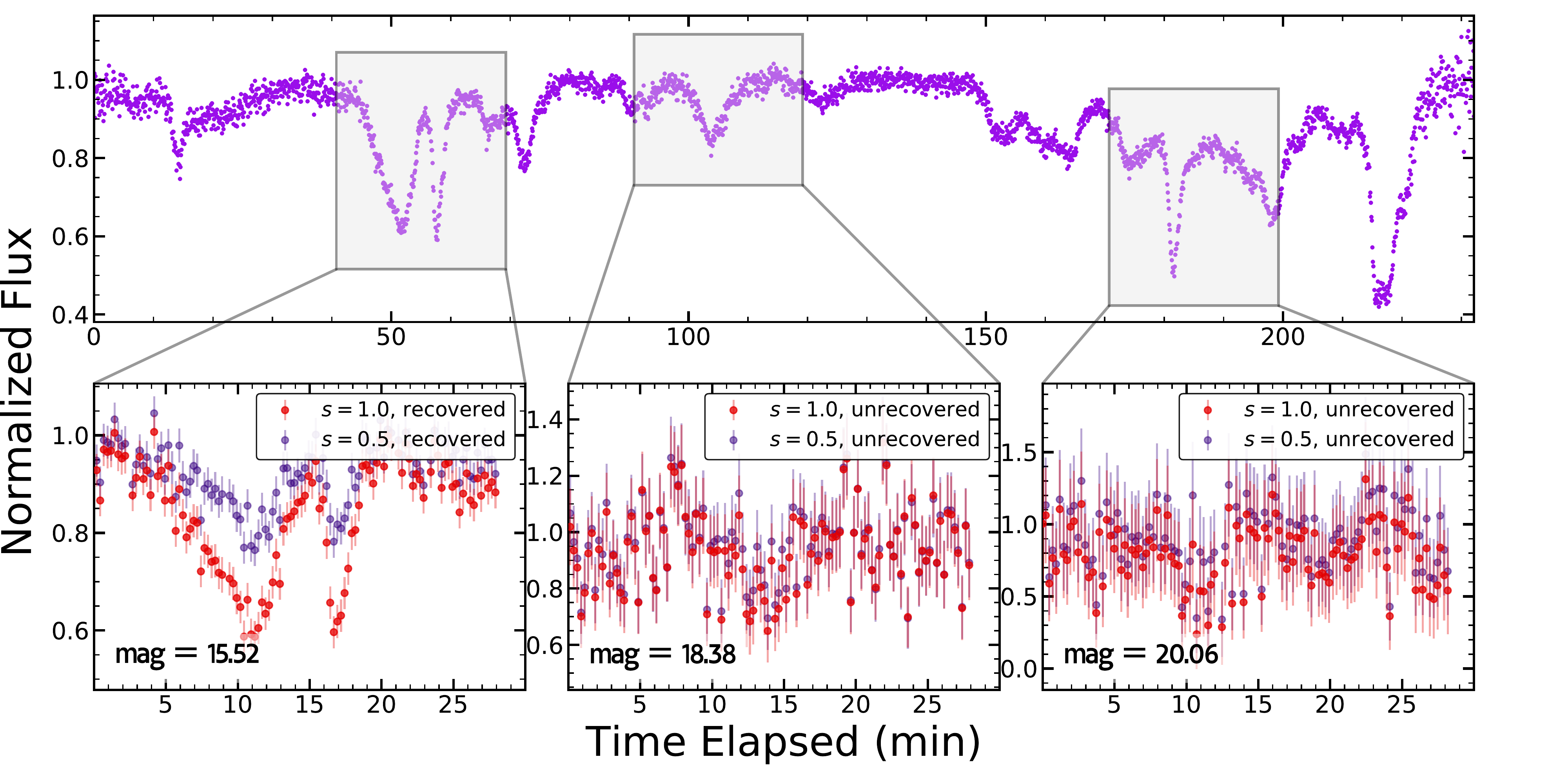}
\caption{Examples of light curve injection for three WDs (PG 0846+558, WD J095143.98$+074956.31$, and US 3608 from left to right) with two scale factors, $s=1.0$ and $s=0.5$. NUV magnitudes are given in the bottom left of each panel.}
\label{fig:compare}
\end{figure*}

\begin{figure*}
\includegraphics[width=\linewidth]{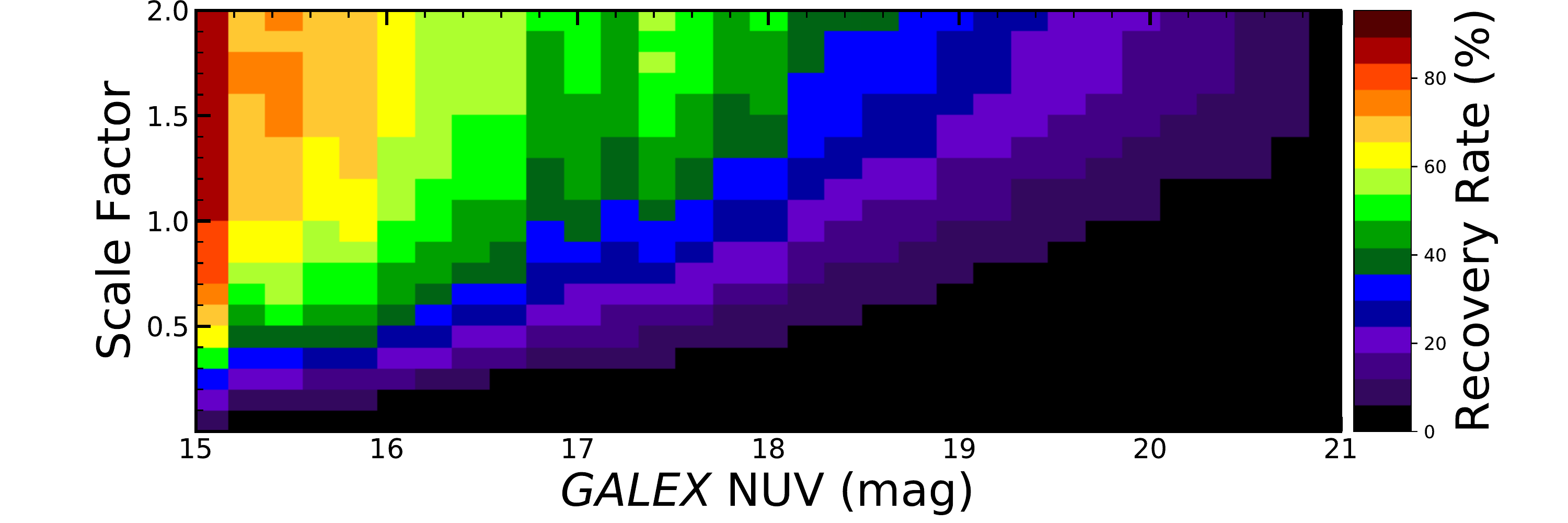}
\caption{Recovery percentage as a function of \textit{GALEX} NUV magnitude and scale factor. As expected, there is a higher recovery rate for larger scales and brighter sources.}
\label{fig:recovery}
\end{figure*}

\begin{figure}
\includegraphics[width=\columnwidth]{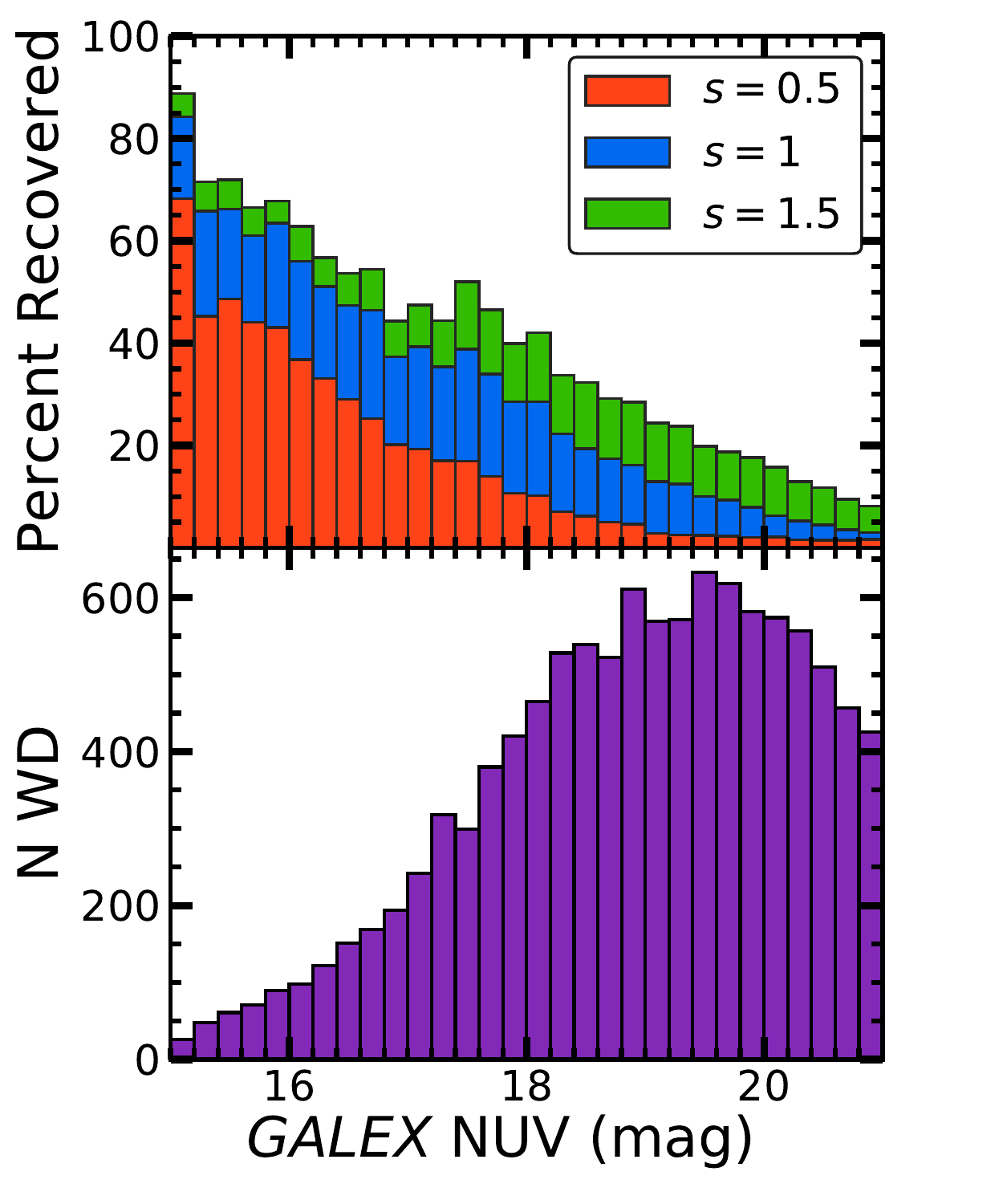}
\caption{Top: Recovery percentage as a function of magnitude for three selected scale factors. Bottom: Number of \textit{GALEX} WDs included in the injection recovery sample for each $0.2$ magnitude bin.}
\label{fig:slices}
\end{figure}

\section{Non-Detection of Planetary Debris Transits} \label{recovery_section}
Despite potential disruption during stellar evolution, a significant fraction of wide planetary systems are expected to survive evolution through the Asymptotic Giant Branch \citep{Mustill12}. Observations of debris disks and IR excesses \citep[e.g.,][]{Barber12,Rocchetto15}, as well as abundances consistent with bulk Earth accreted material \citep[e.g.,][]{Gaensicke12}, provide strong evidence of rocky debris around WDs. Yet direct evidence of transiting planetesimals/asteroids has only been found in WD 1145+017 \citep{Vanderburg15,Gansicke16,Izquierdo18,Karjalainen19}. This source has multiple orbiting bodies with trails of debris, resulting in multiple broad periodogram peaks. Using 1148 K2 WDs, the occurrence rate of transiting material was estimated to be $\sim12\%$ \citep{vanSluijs18}. Though we expect our methodology described in \S\ref{ranking} to detect transiting objects, time-series photometry of \textit{GALEX} WDs does not reveal any such sources. To test recovery of transiting debris, we perform injections of transits from the WD 1145+017 light curve \citep[][Figure 1]{Gansicke16}.

\subsection{Source Selection}
From our initial sample of \surveysize WDs, we include only \textit{GALEX} observations with $>500$ seconds of exposure. This cutoff is based on our detected WD variables, which all have exposure of $>700$ seconds. We also exclude any WDs with rank in the top 5\% of sources, as calculated in \S\ref{ranking}. Additionally, we do not consider any WDs included in the K2 study from \cite{vanSluijs18}, allowing a comparison between independent samples. Since many bright sources are overexposed in the \textit{GALEX} aperture, we limit the brightness at $15$ \textit{GALEX} NUV magnitude. These cuts leave $15034$ WDs in our injection-recovery sample.

To test recovery, we inject portions of the 3.9 hour optical light curve of WD 1145+017 presented in \cite{Gansicke16} and shown in the top panel of Figure \ref{fig:compare}. This light curve has a large range of transit depths and durations, as well as periods of no variability. Since our \textit{GALEX} observations have maximum exposures of 30 minutes, we inject a wide variety of possible transits.

In the first step of the recovery procedure, we iterate through NUV magnitude bins of width $0.2$ mag. A random WD in the magnitude bin is selected from the trimmed sample. If the selected WD has more than one \textit{GALEX} observation, we randomly select a single observation from the set. A segment of the optical light curve is then selected at random, matching the duration of the \textit{GALEX} observation. 

After selecting an optical light curve segment to inject, we multiply the optical flux by a scale factor from 0--2, also chosen at random. If we select random 30-minute segments of the WD 1145+017 LC, 90\% have a depth $>10\%$ in flux, $\sim45\%$ have a depth $>25\%$, and $\sim8\%$ have a depth $>50\%$. The inclusion of a scale factor allows for a wider sampling of possible transit depths, while still remaining true to the distinct patterns of variability observed in WD 1145+017.

Finally, using the methodology described in \S\ref{ranking}, we compute the rank for this observation after injection. Though we had visually inspected the top 10\% of ranked sources in our variability search, we conservatively consider an injected source recovered if the new rank falls within the top 5\% of ranks computed for all WDs. 

Figure \ref{fig:compare} plots an example of injections at three magnitudes and two scale factors. The recovery of a source is not only dependent on the section of the WD 1145+017 optical light curve selected and the scale factor used, but also the magnitude of the source. For fainter sources, the increased scatter masks out injected transits, limiting recovery potential.

\subsection{Recovery Results} \label{RecoveryResults}
We run the injection/recovery routine with \SI{2e5} iterations per 0.2 magnitude bin. With a scale factor bin size of 0.1, there are $\sim10^4$ iterations at each magnitude and scale factor bin. Figure \ref{fig:recovery} plots the recovery percentage for each bin. As expected, recovery percentage is higher for larger scales and brighter sources. Figure \ref{fig:slices} plots the recovery percentage at three selected scale factors and a histogram of source magnitudes. 

Due to the low number of bright sources, there are larger fluctuations in recovery percentage between adjacent magnitude bins of $\lesssim17.5$ NUV mag. Overexposure and higher flag counts also cause more variation in the quality of these bright \textit{GALEX} observations. Comparison of the three sources in Figure \ref{fig:compare} demonstrates a greater change in noise between the 15.52 and 18.38 mag sources then the 18.38 and 20.06 mag sources.

\begin{figure}
\includegraphics[width=\columnwidth]{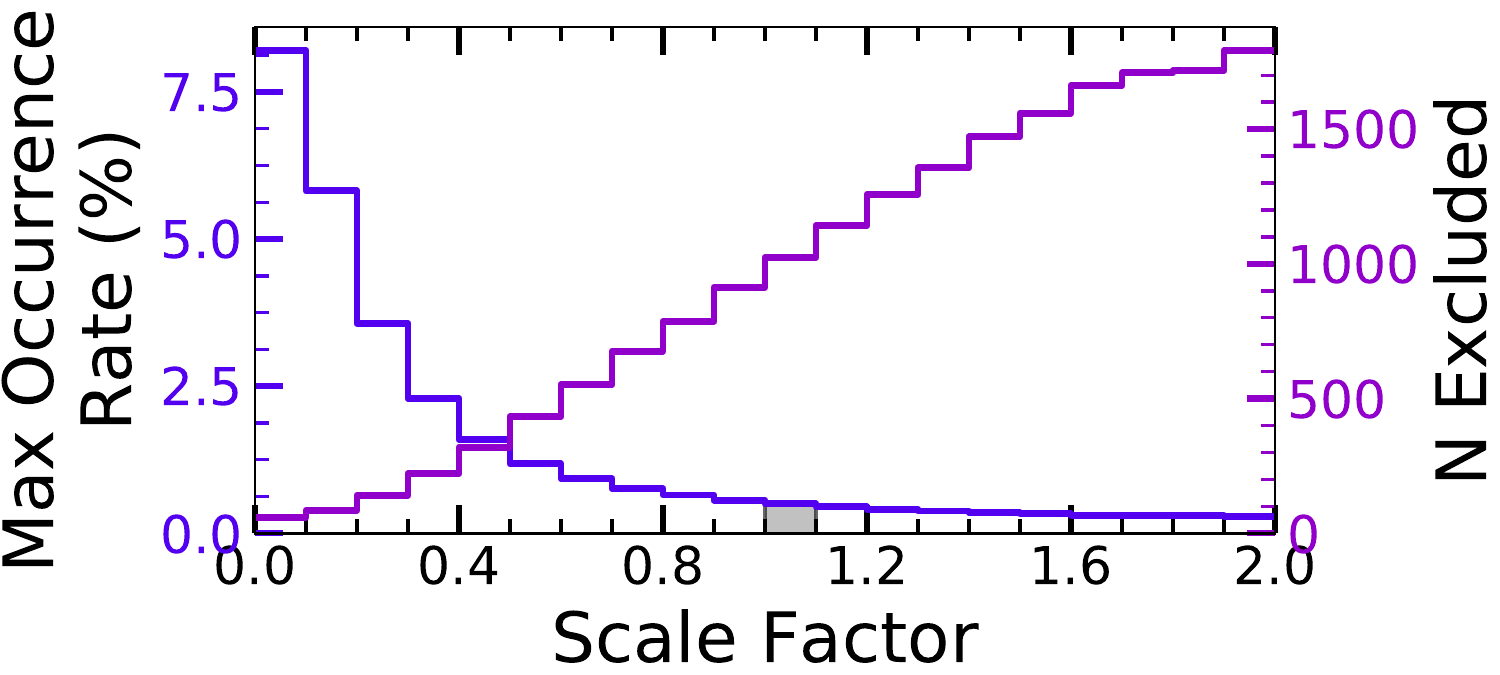}
\caption{Maximum $3\sigma$ occurrence rates of transiting debris calculated for each scale factor. The number of sources excluded from hosting transiting sources at each scale factor is given along the right axis calculated using Equation \ref{eq:n_excluded}. We take the maximum occurrence rate, 0.5\%, to be the $3\sigma$ upper limit at a scale factor of 1.0 (\S\ref{RecoveryResults}).}
\label{fig:occurrence}
\end{figure}

To estimate the number of WDs for which we can rule out transiting debris, we consider the number of WDs in a given magnitude bin, shown in the bottom panel of Figure \ref{fig:slices}, multiplied by the recovery rate at a chosen scale factor:

\begin{equation} \label{eq:n_excluded}
N_{\text{excluded,}s} = \sum_{m=15}^{21} N_{m}\times R_{m, s}
\end{equation}
where $m$ is the \textit{GALEX} NUV magnitude, $N_{m}$ is the number of sources used for injection in the $m^{\text{th}}$ magnitude bin (Figure \ref{fig:slices}), and $R_{m,s}$ is the recovery fraction calculated at the corresponding magnitude and scale factor (Figure \ref{fig:occurrence}). This value represents the estimated number of WDs excluded from hosting detectable transits. To place an upper limit on the occurrence rate of transiting debris, we consider a binomial population where a detected transit is considered a ``success". Since no transiting objects are present in our sample, we use the equal-tailed Jeffreys Interval to estimate a $3\sigma$ upper limit of zero successes in $N_{\text{excluded}}$ trials. 

Figure \ref{fig:occurrence} plots $N_{\text{excluded}}$ and the resulting $3\sigma$ occurrence upper limits as a function of the scale factor. Under the assumption the material is opaque, we calculate our $3\sigma$ upper limits at a scale factor of $s=1.0$. We find 1024 WDs are ruled out from having transiting debris in our study, leading to a $3\sigma$ occurrence rate upper limit on transiting debris around WDs of 0.5\%. We emphasize that this limit does not describe the occurrence rate of debris, but the rate of \textit{transiting} debris given non-detections in \textit{GALEX} photometry. The full results of the occurrence rate calculation for all scale factors are provided as supplementary material. 

This occurrence rate does not take into account recovery of different body sizes or orbital separations, only the detection of transiting debris analogous to WD 1145+017. Using 1148 WDs from K2, combined with an estimated transit probability of 2\% \citep{Vanderburg15}, \cite{vanSluijs18} reported a 12\% occurrence rate with 95\% confidence bounds spanning 1-45\%. Taking the same consideration into account, our calculated occurrence rate is consistent with that of \cite{vanSluijs18}. The true occurrence rate of these systems also depends on the observed inclination, modifying our upper limit by a factor of $\cos(i)$, as well as the UV-to-optical scale ratio of the transit depths. Even if the transits are half as deep in the UV ($s=0.5$), our $3\sigma$ occurrence rate upper limit is still $<$2\% (see Figure~\ref{fig:occurrence}).

\section{Conclusions} \label{conclusion}

We present the largest UV photometric WD survey to-date in a search for new variable WDs. Our sample of \surveysize WDs comes from four separate catalogs, which was trimmed based on \textit{GALEX} exposure time.  We demonstrate the application of short baseline \textit{GALEX} observations for the detection of photometric variability. 

Applying a system of weighted metrics results in the detection of 37 new DA pulsators and four new DB pulsators. We provide the first eclipse detection of eight eclipsing WD binary systems. Appendix \ref{appendixLC} plots light curves of all new pulsators and Figure \ref{fig:EclipseTower} plots the nine eclipsing sources. Using \textit{Gaia} color information, we classify these pulsators as either DAV or DBV using a CMD (Figure \ref{fig:cmd}). Included in our new pulsators is the detection of a new massive pulsator, WD J212402.04$-$600100.05 (\textit{Gaia} $G$=18.0 mag), of which we are currently conducting follow-up observations.

We detect variability in sources up to 20th $G$-band apparent magnitude, with a median of $\sim18^{\rm{th}}$ magnitude. \textit{GALEX}/{\sc gPhoton} are more successful in searching for fainter pulsators, as compared to the known population \citep{Bognar16}. This is partially due to oversaturation of bright sources in the \textit{GALEX} aperture, preventing the detection of some known pulsators. For sources where observations were taken in FUV and NUV bands, there is consistent pulsation variability timescales in each. By considering multiple metrics to identify variability, we are also able to identify pulsations and eclipses when only one band is available.  

While this study provides a comprehensive analysis of \textit{GALEX} WD photometry, there are almost certainly additional low amplitude variability candidates to be found around fainter sources. This study does not provide any estimates on detectability or occurrence rates of WD pulsators. There are also some sources where detection is limited by instrumental effects, most notably the telescope dither and oversaturation of the \textit{GALEX} aperture.

Though our method of variability selection did not return any candidates for transiting exoplanet debris, detection of these sources could still be possible with time-series photometry of \textit{GALEX} observations. For sources with multiple visits, it is feasible to detect multi-periodic transits, similar to those detected by \cite{Vanderburg15}. We perform synthetic injections of the WD-1145+017 light curve from Figure 1 of \cite{Gansicke16} and use the system of weighted metrics to assess recovery. We find the maximum $3\sigma$ occurrence rate of WD 1145+017-like transiting objects to be 0.5\%.

Both pulsators and eclipsing binary sources presented here are potential targets for longer-baseline studies. Timing of detected pulsation modes can be used for asteroseismology. Detection and classification of these sources is a continuation in probing WD interiors to better understand the final stage of stellar evolution for the majority of stars. \\

\section*{Acknowledgements}
The authors would like to thank Daniel Huber for his suggestions on variability identification and Mark Seibert for his suggestions on \textit{GALEX} data usage. We thank Ashley Chontos, Sam Grunblatt, and Siyi Xu for useful discussions and improvements on the manuscript. Many thanks to Boris G\"{a}nsicke for providing the WD 1145+017 light curve. We also thank Scott Fleming and Chase Million for assistance with {\sc gPhoton} data retrieval and troubleshooting. 

DMR acknowledges support from Research Experience for Undergraduate program at the Institute for Astronomy, University of Hawaii-Manoa funded through NSF grant AST-1560413 and would like to thank the Institute for Astronomy, Eugene Magnier, \& Nader Haghighipour for their hospitality. MAT acknowledges support from the United States Department of Energy through the Computational Sciences Graduate Fellowship (DOE CSGF). Support for JJH was provided by NASA through Hubble Fellowship grant \#HST-HF2-51357.001-A, awarded by the Space Telescope Science Institute, which is operated by the Association of Universities for Research in Astronomy, Incorporated, under NASA contract NAS5-26555.

Data presented in this paper was obtained from the Mikulski Archive for Space Telescopes (MAST). STScI is operated by the Association of Universities for Research in Astronomy, Inc., under NASA contract NAS5-26555. Support for MAST for non-HST data is provided by the NASA Office of Space Science via grant NNX09AF08G and by other grants and contracts.

This research has made use of the SIMBAD database, operated at CDS, Strasbourg, France \citep{Wegner00}. This research has made use of the VizieR catalogue access tool, CDS, Strasbourg, France \citep{Oschsenbein00}.

This work has made use of data from the European Space Agency (ESA) mission {\it Gaia} (\url{https://www.cosmos.esa.int/gaia}), processed by the {\it Gaia} Data Processing and Analysis Consortium (DPAC, \url{https://www.cosmos.esa.int/web/gaia/dpac/consortium}). Funding for the DPAC has been provided by national institutions, in particular the institutions participating in the {\it Gaia} Multilateral Agreement.\\ \indent The Pan-STARRS1 Surveys (PS1) and the PS1 public science archive have been made possible through contributions by the Institute for Astronomy, the University of Hawaii, the Pan-STARRS Project Office, the Max-Planck Society and its participating institutes, the Max Planck Institute for Astronomy, Heidelberg and the Max Planck Institute for Extraterrestrial Physics, Garching, The Johns Hopkins University, Durham University, the University of Edinburgh, the Queen's University Belfast, the Harvard-Smithsonian Center for Astrophysics, the Las Cumbres Observatory Global Telescope Network Incorporated, the National Central University of Taiwan, the Space Telescope Science Institute, the National Aeronautics and Space Administration under Grant No. NNX08AR22G issued through the Planetary Science Division of the NASA Science Mission Directorate, the National Science Foundation Grant No. AST-1238877, the University of Maryland, Eotvos Lorand University (ELTE), the Los Alamos National Laboratory, and the Gordon and Betty Moore Foundation.



\bibliographystyle{mnras}
\bibliography{DomsBib}



\appendix
\section{New WD Pulsator Spectra} \label{appendixSpectra}

Using the SuperNova Integral Field spectrograph \citep[SNIFS]{Lantz04} on the University of Hawaii 88-inch telescope (UH88), we present spectroscopic observations for two of the new pulsators. These spectra were taken to confirm our classification of the new DA/B pulsators via the Gaia CMD (Figure \ref{fig:cmd}). The spectra cover roughly 3200--9000\AA, excluding the dichroic crossover ($\sim$5000--5200\AA). 

\begin{figure*}
	\centering
    \includegraphics[width=\linewidth, keepaspectratio]{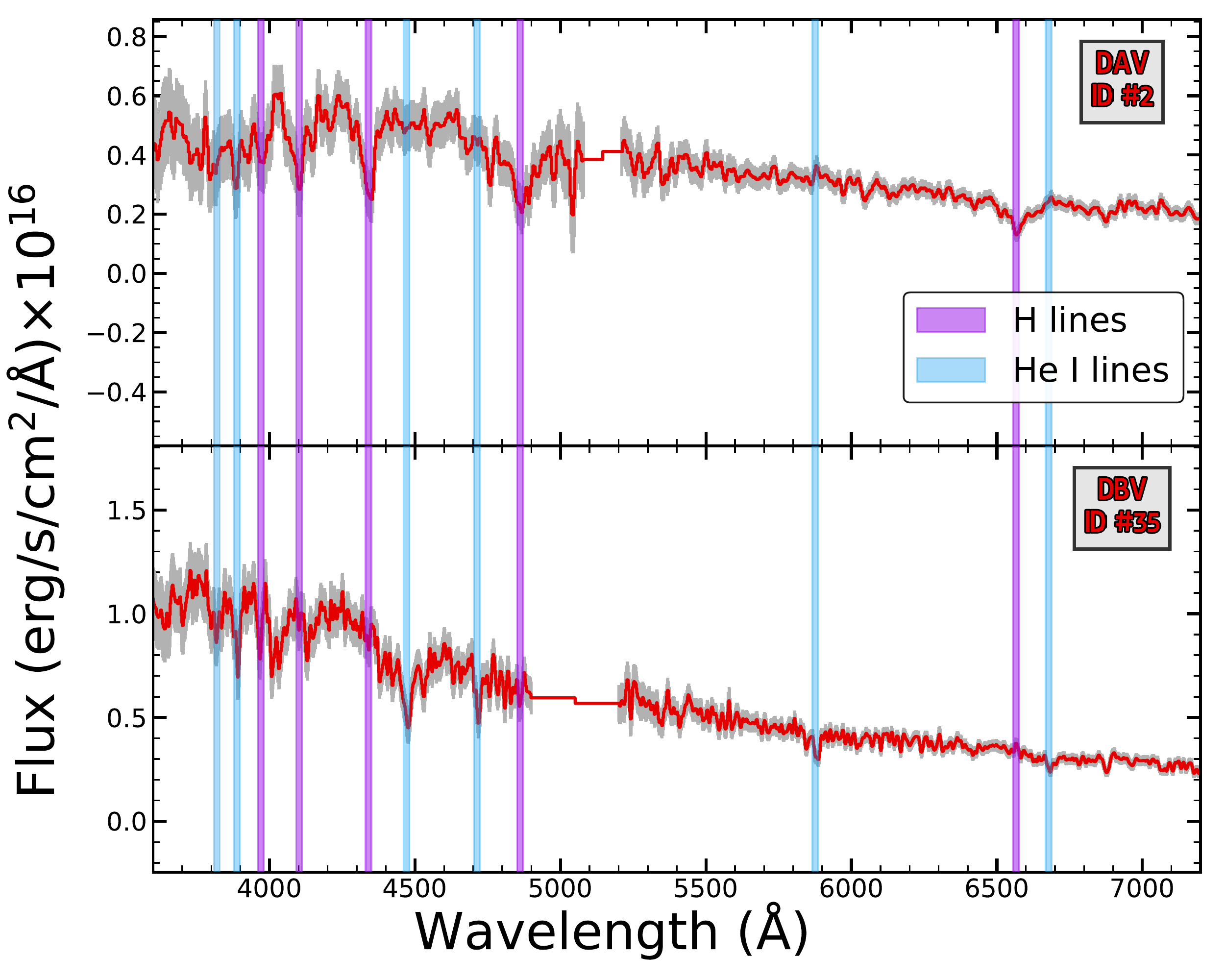}
    \caption{UH88/SNIFS spectra for two new pulsators with error bars in gray. Top: DAV pulsator with H absorption lines. Bottom: DBV pulsator with He absorption lines. These spectra confirm our classifications based on the Gaia CMD in Figure~\ref{fig:cmd}.}
    \label{fig:newspectra}
\end{figure*}

The spectrum for WD J003116.51$+$474828.39 (ID\#2), provided in the top panel of Figure \ref{fig:newspectra}, was taken on MJD 58301.584757 with an exposure time of 1820 seconds at an airmass of 1.272. The spectrum exhibits broad H absorption lines, notably at 4340, 4861, and 6563\AA, confirming our initial DAV classification.

The spectrum for WD J132952.63$+$392150.8 (ID\#35), provided in the bottom panel of Figure \ref{fig:newspectra}, was taken on MJD 58321.301690 with an exposure time of 1520 seconds at an airmass of 1.480. The spectrum exhibits He I features at 4471, 4712, 5876, and 6678\AA, again confirming our DBV classification.

\section{Pulsator Light Curves} \label{appendixLC}
We create UV lightcurves for all detected pulsators presented in this study in Figure \ref{fig:LCappendix1}. When available, NUV and FUV data are plotted, in red and blue, respectively. For cases where there are more than one visit displaying variability, we plot the highest ranked light curve. As expected \citep{Kepler2000},
we see larger amplitude variability in the FUV. 
\begin{figure*}
    \centering
    \includegraphics[width=\linewidth, keepaspectratio]{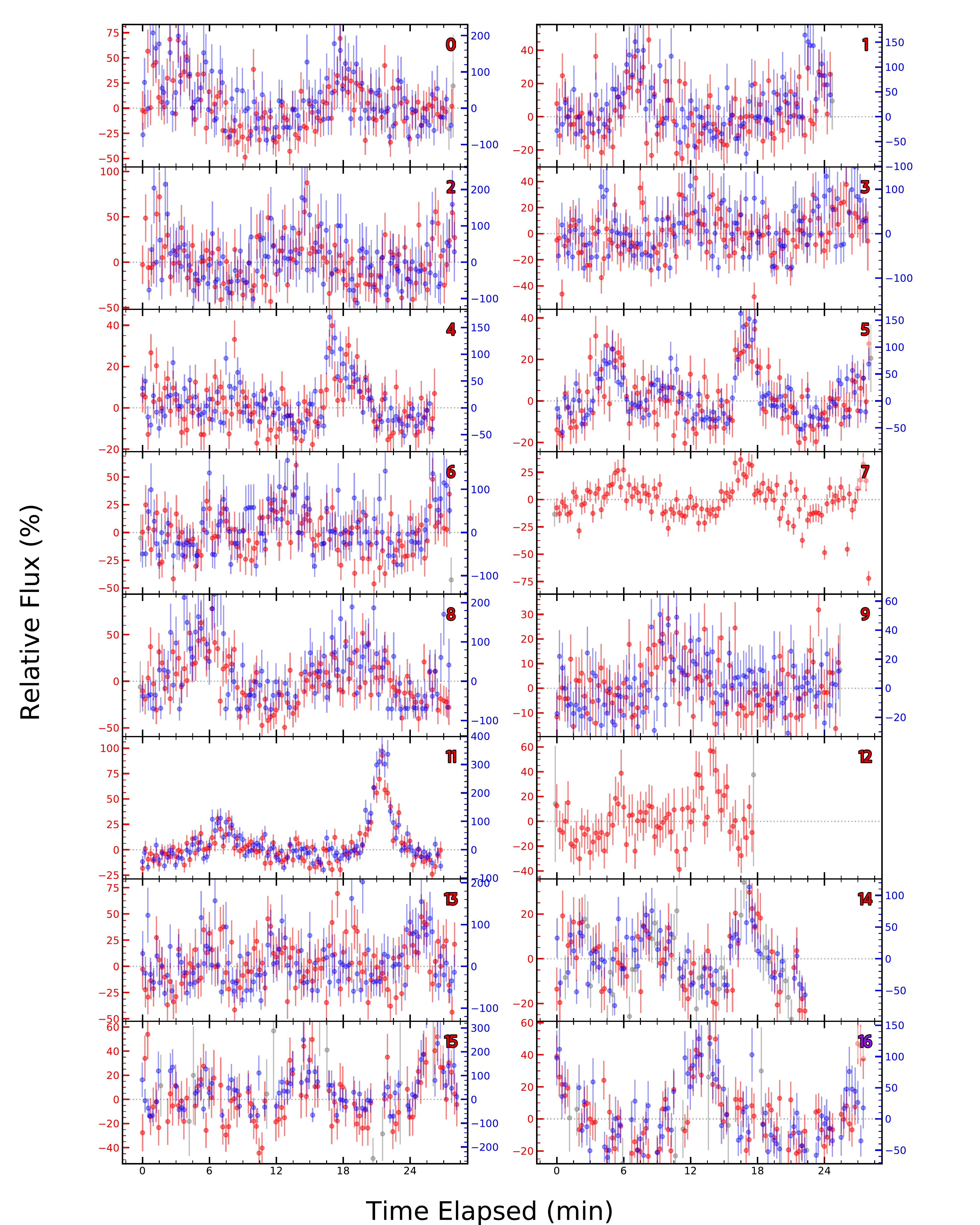}
    \caption{Light curves for all pulsators detected in our survey. ID numbers from Table \ref{tab:allsources} are given in top right of each panel, with new pulsators in red and known pulsators in purple. NUV data is plotted in red, FUV in blue, and flagged points in gray.}
    \label{fig:LCappendix1}
\end{figure*}

\begin{figure*}
    \centering
    \includegraphics[width=\linewidth, keepaspectratio]{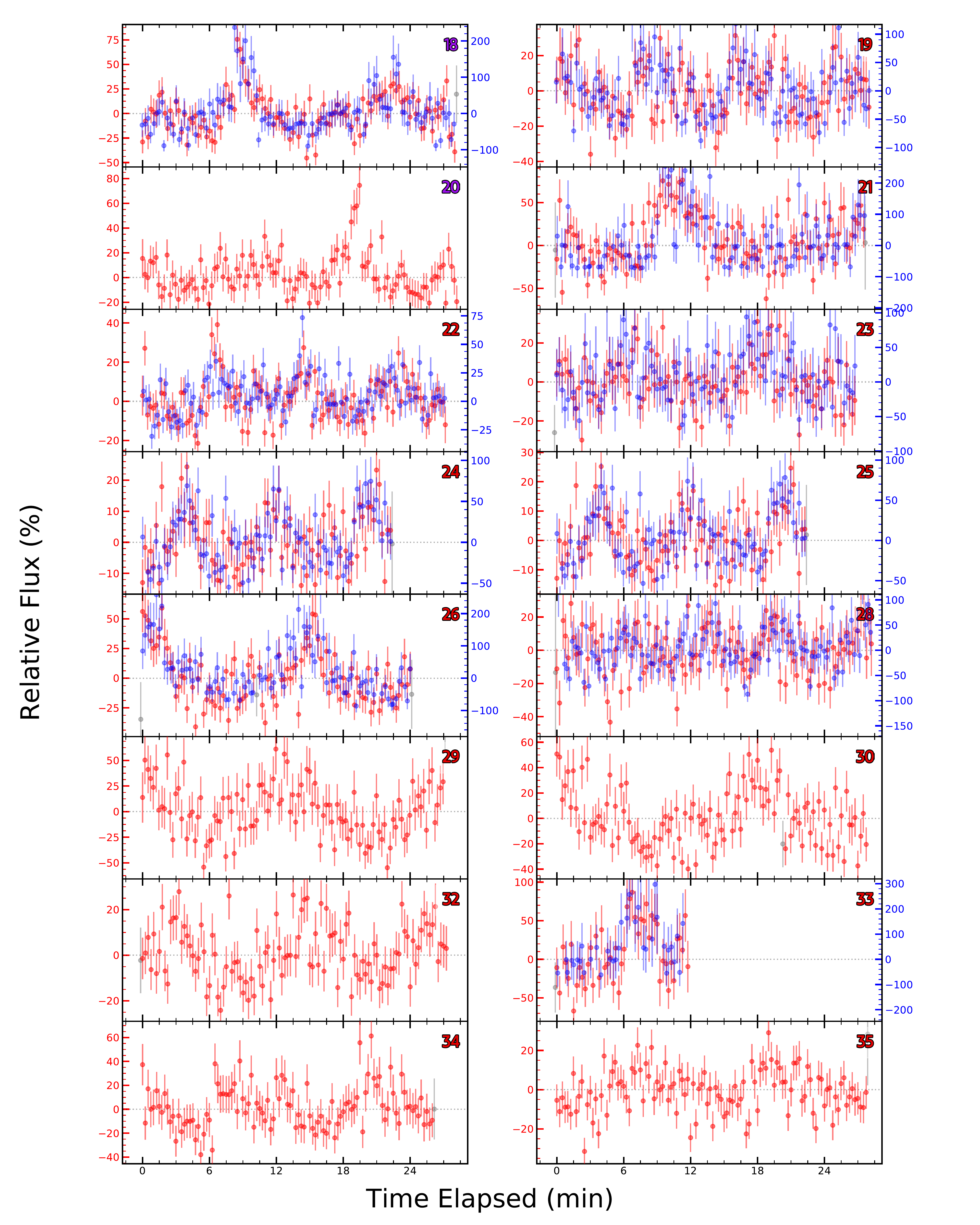}
    \caption{Continuation of Figure \ref{fig:LCappendix1}}
    \label{fig:LCappendix2}
\end{figure*}

\begin{figure*}
    \centering
    \includegraphics[width=\linewidth, keepaspectratio]{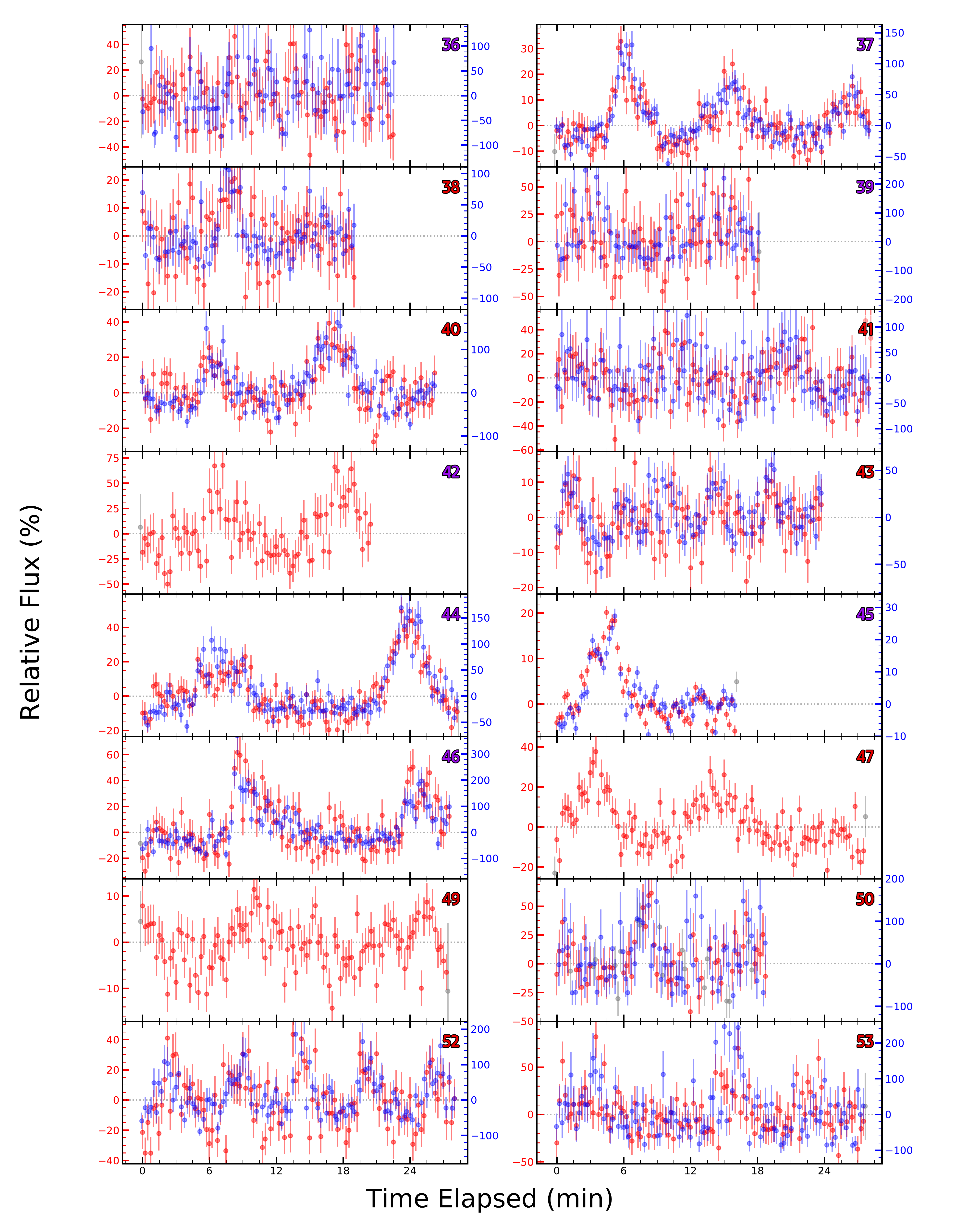}
    \caption{Continuation of Figure \ref{fig:LCappendix1}}
    \label{fig:LCappendix3}
\end{figure*}

\begin{figure*}
    \centering
    \includegraphics[width=\linewidth, keepaspectratio]{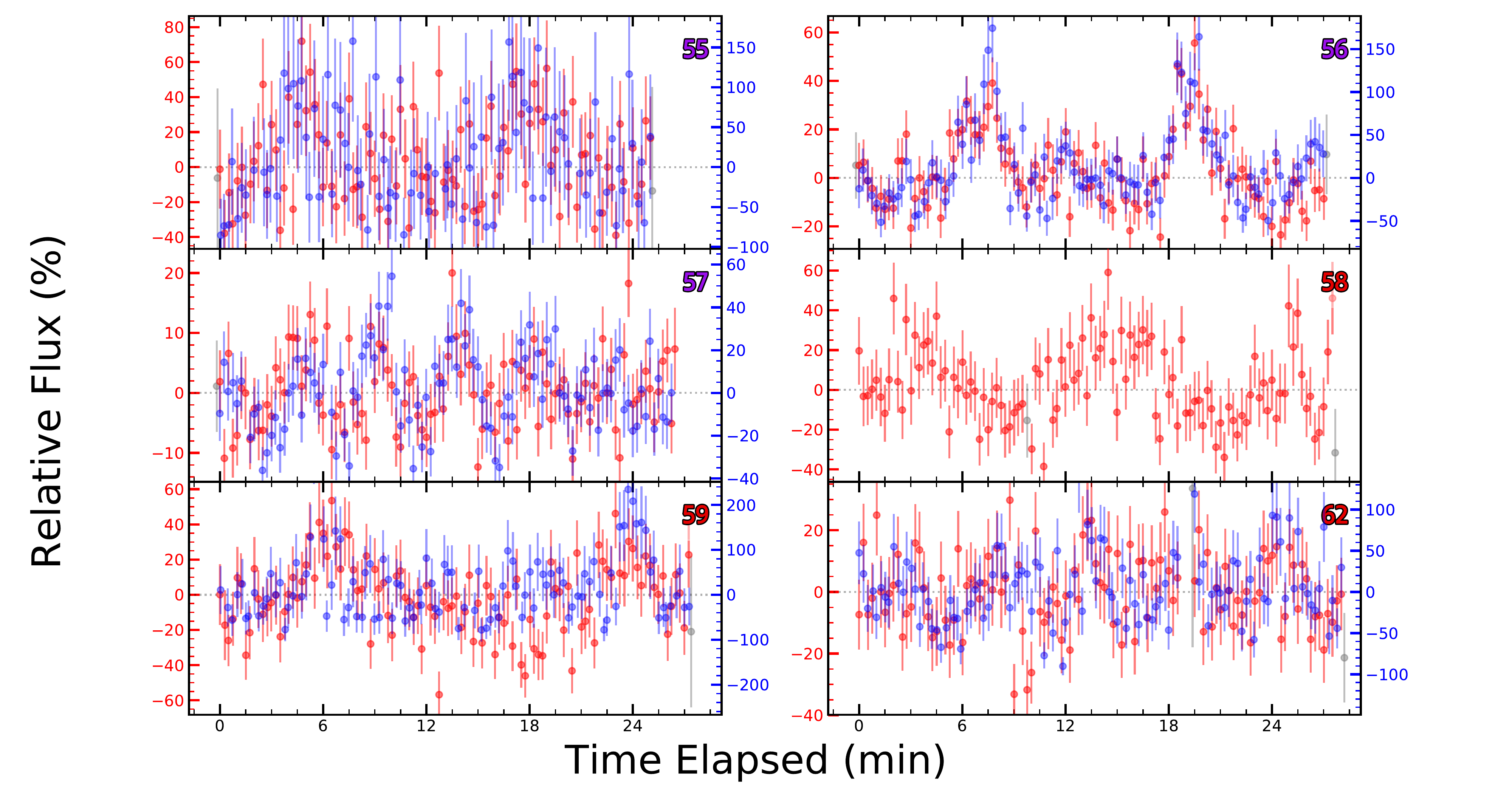}
    \caption{Continuation of Figure \ref{fig:LCappendix1}}
    \label{fig:LCappendix4}
\end{figure*}


\bsp	
\label{lastpage}
\end{document}